\documentclass[11pt,a4paper]{article}

\usepackage{fullpage}
\usepackage{times}

\usepackage[T1]{fontenc}
\usepackage{amsmath}%
\allowdisplaybreaks
\usepackage{amssymb}%
\usepackage{epsfig}%
\usepackage{amsthm}%
\usepackage{dsfont}%
\usepackage{natbib}
\usepackage{url}%
\usepackage{mathrsfs}%
\usepackage{braket}%
\usepackage{accents}
\usepackage{caption}
\usepackage{subcaption}
\usepackage{color}
\usepackage{appendix}
\usepackage{cases}
\usepackage{enumerate}
\usepackage{enumitem}
\usepackage{mathtools}
\mathtoolsset{showonlyrefs} 
\usepackage{graphicx}

\advance\voffset by 0pt
\advance\hoffset by 0pt
\renewenvironment{proof}[1][Proof]{\noindent\textbf{#1.} }{\ \rule{0.5em}{0.5em}}

\theoremstyle{plain}%
\newtheorem{theorem}{Theorem}
\newtheorem{lemma}{Lemma}

\newtheorem{cor}{Corollary}

\newtheorem{proposition}{Proposition}
\newtheorem{remark}{Remark}

\theoremstyle{definition}%
\newtheorem{definition}{Definition}
\newtheorem{assumption}{Assumption}

\renewenvironment{appendix}%
{\setcounter{section}{0}%
	\numberwithin{equation}{section}%
}%
{}%

\newcommand{\sr}{\mathds{R}}%
\newcommand{\Ep}{\textup{\textsf{E}}}%
\newcommand{\Pp}{\textsf{P}}%
\newcommand{\keyword}[1]{\noindent\textit{Keywords:} #1\\}%
\newcommand{\pro}[1]{\left(#1_t\right)_{t\geq 0}}%
\newcommand{\rd}{\mathrm{d}}%

\newcommand{\cF}{{\mathscr F}}%
\newcommand{\cD}{{\mathscr D}}%
\newcommand{\cP}{{\mathscr P}}%
\newcommand{\cL}{{\mathscr L}}%
\newcommand{\cQ}{{\mathscr Q}}%

\newcommand{\bF}{{\mathbf F}}%
\newcommand{\xL}{{\underline{x}}}%
\newcommand{\xU}{{\overline{x}}}%
\newcommand{\xS}{{x^{\ast}}}%
\newcommand{\vC}{{\check{\varphi}}}%
\newcommand{\vH}{{\hat{\varphi}}}%
\newcommand{\ts}{\tau^\ast}%
\newcommand{\es}{\eta^\ast}%
\newcommand{\ps}{\phi^\ast}%
\newcommand{\cc}{\check c }%
\newcommand{\ch}{\hat c  }%

\newcommand{\dt}{\mathrm{d}}%

\usepackage{todonotes}
\usepackage{soul}

\usepackage{hyperref}
\hypersetup{%
	colorlinks=true, 
	linkcolor=blue!80!black, 
	citecolor=green!60!black, 
	filecolor=magenta,     
}

\title{Singular Control in a Cash Management Model with Ambiguity}
\author{Arnon Archankul\thanks{Department of Mathematics, University of York, United Kingdom. \ttfamily{arnon.archankul@york.ac.uk}} \hspace{0.5mm} Giorgio Ferrari\thanks{Center for Mathematical Economics (IMW), Bielefeld University, Germany. \ttfamily{giorgio.ferrari@uni-bielefeld.de}} \hspace{0.5mm} Tobias Hellmann\thanks{HARTING Technology Group, Espelkamp, Germany. \ttfamily{tobias.hellmann@harting.com}} \hspace{0.5mm} and Jacco J.J. Thijssen\thanks{Department of Mathematics, University of York, United Kingdom. \ttfamily{jacco.thijssen@york.ac.uk}}}

\begin{document}
	\maketitle
	
	\begin{abstract}
		
		We consider a singular control model of cash reserve management, driven by a diffusion under ambiguity. The manager is assumed to have maxmin preferences over a set of priors characterized by $\kappa$-ignorance. A verification theorem is established to determine the firm's cost function and the optimal cash policy; the latter taking the form of a control barrier policy. In a model driven by arithmetic Brownian motion, we use Dynkin games to show that an increase in ambiguity leads to higher expected costs under the worst-case prior and a narrower inaction region. The latter effect can be used to provide an ambiguity-driven explanation for observed cash management behavior. Our findings can be applied to broader applications of singular control in managing inventories under ambiguity.

		\keyword{Singular control, Ambiguity, Inventory models}%
	\end{abstract}%
	
	
	\section{Introduction}\label{sec:intro}
	
	An important question in corporate finance is that of optimal cash management. On the one hand, firms require cash to finance the firm as a going concern. On the other hand, shareholders require dividend payouts as a reward for providing capital. The seminal contribution by \cite{Shiryaev95} uses a stochastic storage models \emph{\`{a} la} \cite{harrison1983instantaneous} to find the optimal size of a firm's cash hoard in the face of stochastically evolving net cash flows. In this paper, we are interested in optimal cash management under ambiguity, i.e., a situation where the manager is not able to reduce the uncertainty over future net cash flows into a single probability measure. We are interested in the interplay between traditional concerns over risk (as measured by, e.g., confidence intervals provided by a given probability measure) and ambiguity (as measured by the ``size'' of the set of probability measures considered by the manager) under the assumption that the manager is ambiguity averse. Furthermore, we demonstrate that our results apply more broadly to singular control of inventories under ambiguity.
	
	Our motivation for incorporating ambiguity into the singular control framework of cash holding stems from  empirical evidence that ambiguity exerts a first-order effect on corporate cash management. A particularly relevant study is that of \citet{breuer2017corporate}, which investigates how firms adjust their cash policies in response to investors' ambiguity aversion. They find that ambiguity averse investors tend to undervalue uncertain future investment opportunities, thereby diminishing the perceived value of holding cash for such purposes. In turn, firm managers respond by actively reducing cash holdings, often through dividend payouts, to align with investor preferences and preserve firm value. This adjustment is not a passive by-product of other financial decisions but, rather, a direct policy response to ambiguity.
	
	While a broader empirical literature links ambiguity to firms’ financial decisions, much of it centers on investment behavior, with cash policy treated as a secondary outcome. For example, studies such as \citet{Neamtiu2014-cv,Goodell2021-fk,Luo2022-zh,Zhao2023-rg} document how macroeconomic ambiguity, arising from such as inflation, GDP volatility, unemployment, or economic policy uncertainty, affects managerial incentives and capital expenditures, with subsequent implications for cash holdings. In these cases, ambiguity affects cash indirectly through its impact on investment timing, implying a second-order effect.
	
	From a theoretical perspective, such dynamics are well captured by models of investment under ambiguity, notably the real options framework developed by \citet{Nishimura2007-it}. However, existing models largely abstract from the possibility that cash itself may be the principal adjustment mechanism to ambiguity. To fill this gap, our paper develops a stochastic control model of cash management under ambiguity, where cash holding responds directly and optimally to ambiguity. This allows us to characterize the first-order role of ambiguity in shaping firms' cash policies, and to explore the implications for corporate financial behavior in uncertain environments.

	In this paper, we develop a framework in which a firm’s decision maker (DM) can dynamically adjust cash holdings, where holding cash is costly and cash adjustments may incur additional costs. We assume that the DM has a \emph{reference prior}, possibly elicited from available data or based on their industry experience, but is \emph{ambiguous} about the true probability measure. The DM then dynamically uses the worst-case prior to determine the optimal cash policy.

	The distinction between uncertainty resulting from randomness governed by a distribution (``known unknowns'') and uncertainty over the correct distribution (``unknown unknowns'') goes back to \citet{knight1921risk}. In his seminal work he refers to the former as \emph{risk} and the latter as \emph{uncertainty} or \emph{ambiguity}. The effect of ambiguity on decision making has been studied extensively, most famously by \cite{El61}. The overwhelming conclusion of the experimental literature is that DM are \emph{ambiguity averse}. In the classical Ellsberg experiment, a DM has to place bets on one of two urns, both with 100 red or blue balls. For the first urn it is known that half the balls are red. For the second urn no such information is available. Since most people are observed to choose bets on the first urn over bets on the second urn, Savage's ``sure thing principle'' is being violated. When compared to previous empirical findings, this experiment can be interpreted as if the firms' manager are confronted with the second Ellsberg urn. This, as a result, prompts them to assign their subjective priors.
	
	Note that the Ellsberg paradox is not really a paradox, because it does not result from a cognitive bias or irrationality. Rather, observed behavior is driven by a lack of information. It is perfectly possible for DMs to make consistent decisions under ambiguity. This has been shown by \cite{GiSch89}, who incorporate an ambiguity aversion axiom into the subjective expected utility framework. They then show that a rational DM acts \emph{as if} she maximizes expected utility over the worst--case prior within a (subjectively chosen) set of priors. 
	This approach has been successfully used in many applications in economics, finance, and operation research.\footnote{See, for example of applications in investment decisions, \citet{Nishimura2007-it, Trojanowska2010-zn, Thijssen2011-ag, Cheng2013-to, Hellmann2018-xs} delve into timing game under ambiguous environment, while \citet{Asano2021-wg,Driouchi2020-zf} incorporate ambiguity into the model by means of technology shocks, and cultural biases, respectively. The works of \citet{Jin2015-op,fouque2016portfolio,Lin2021-fo} apply ambiguity to portfolio management. 
	For the broader theory of ambiguity in volatility and interest rate in asset pricing, we refer to \citet{Epstein2013-fj} and \citet{Lin2021-fo}, respectively. For the literature related to decision making under \emph{smooth ambiguity}, which is another approach to model of ambiguity introduced by \citet{KlMaMu05}, we refer, for example, to the work of \citet{Hansen2018-yc,Hansen2022-ob,Balter2021-te,Balter2020-mb,Suzuki2018-nj,borgonovo2015decision}, among other notable authors. }
	
	Our contribution is to apply the maxmin multiple prior model to a singular control model of optimal storage inventory, with an application to a firm's cash management. On a regular basis, firms are faced with  operational costs (e.g. rent, capital stock, labor's wage, etc.) that have to be settled promptly with reserved cash. The fact that this cash generates no (or low) return means holding it results in an opportunity loss, which can be interpreted as a holding cost as it could potentially be used for income-generating activities, such as investments or paying out dividends. Therefore, excessive cash holding is undesirable. On the other hand, having a shortage of cash reserves results in a delay of cost settlement, which often incurs a penalty fee or credit loss. Therefore, the firm has an incentive to inject some amount of cash into the system. This could, for example, be done by selling some assets or issuing bonds. These two circumstances create a trade-off that suggests the existence of target level of cash. In a model where cash adjustments are costly, we show that there exists an optimal \emph{control band policy}, where the firm keeps its cash hoard between an upper and lower bound. The cash reserve problem was first addressed in the literature by \citet{Baumol1952-rz} and \citet{Tobin1956-zd}, who studied the cash balance problem under the assumption that demand is deterministic, which is far from realistic. The stochastic treatment was later established under a discrete-time (Markov chain) framework by e.g. \citet{Eppen1969-js}. A more general approach for storage systems in continuous time, in particular, with demand driven by Brownian motion, has been developed over the past decades. \citet{Bather1966-wi,vial1972continuous,Constantinides1976-rf,Michael_HARRISON1978-dq,harrison1983instantaneous,Dai2013-yz,Dai2013-xb} and many others are among the notable authors. To get an overview of the related papers, we refer the reader to \citet{Michael_Harrison2013-fe}. 
	
	While this is no different from a standard model under risk, ambiguity does bring some new aspects to the comparative statics of the optimal policy. For example, as in the standard model without ambiguity, the higher the risk, the higher the long-term discounted cost of cash. Ambiguity amplifies this effect, even though an increase in the degree of ambiguity leads the manager to exert control \emph{earlier}. This is in contrast to the risk-only model, where an increase in risk leads the manager to exert control \emph{later}. 
	
	The reason for this result is that a more ambiguous DM expects the cash level to increase (when positive) or decrease (when negative) more rapidly (in expectation) than a less ambiguous DM. Since holding costs are increasing in the absolute value of the cash hoard, a more ambiguous DM will, thus, exert control sooner. In our model, this behavior is not due to irrationality, but an aspect of the uncertain environment that the manager faces.

	One of the first papers to axiomatize  ambiguity is \citet{GiSch89}. They model ambiguity as a set of priors, among which the DM (subjectively) selects the one that maximises the DM's expected utility. Under an axiom of ambiguity aversion, the prior that is chosen is called the \textit{worse-case prior}, which captures the intuition that an ambiguity-averse  DM is cautious about their beliefs and heavily weighs the possibility of undesirable consequences of their decision. The Gilboa--Schmeidler criterion has become known as \textit{maxmin utility}. However, the Gilboa--Schmeidler framework is a static one and is, thus, insufficient for dealing with situations where the worst-case prior might change over time. An inter-temporal version was proposed by \citet{Epstein1994-bu} in discrete time and by \citet{ChEp02} in continuous time. In these models, the worst-case prior is updated in a Bellman principle-like  one-step-ahead procedure. In order to make this work, attention is restricted to sets of priors that are called \emph{strongly rectangular}. We use the  \citet{ChEp02} approach to modeling multiple priors.
	
	In fact, we use a stronger assumption, also introduced in \citet{ChEp02}, and assume that ambiguity takes the form of \emph{$\kappa$-ignorance}. That is to say, the DM has a reference probability, which is distorted through a density generator. The density generator is assumed to take values in an interval $[-\kappa,\kappa]$, so that the reference prior together with the parameter $\kappa$ determines the set of priors that is considered by the DM. While restrictive, an advantage of this approach is that the degree of ambiguity can be seen to be measured by $\kappa$. 
	
	Importantly, in our model the worst-case prior is not constant but varies over time, depending on the evolution of the actual amount of cash currently held. This unusual feature has been observed by \cite{Cheng2013-to} in the context of pricing a straddle option and \citet{Hellmann2018-xs}. The latter paper models a timing game between two firms contemplating an investment opportunity under ambiguity and show that ambiguity aversion has two effects: ambiguity over future demand (fear of the market), as in the standard literature, but also ambiguity over the other firm's investment decision (fear of the competitor). These have opposite effects on what constitutes the worst-case prior. It turns out there is a threshold to distinguish which type of ambiguity dominates through time. Our model has a similar feature in that control costs are incurred whether at the upper or lower barrier. The worst-case priors at each of these barriers are opposite and this leads, in turn, to the existence of a threshold somewhere in the inaction region (endogenously determined) that separates two regions where different measures constitute the worst-case prior.  
	
	The closest contribution to our work is \citet{chakraborty2023optimal} in which a one-side singular control of a firm's dividend payout policy is considered under ambiguity. They assume, in addition to the classical singular control, that there is a penalty cost associated with a change of measure, which is determined by the Kullback-Leibler divergence. The use of Kullback-Leibler divergence as a model for multiple priors is well established in the literature on robust control; see, e.g., \citet{anderson2003quartet,maenhout2004robust,Hansen2006-tt,Hansen2010-ko,Hansen2011-ix,Hansen2018-yc,Hansen2022-ob,Ferrari2022-sf} and references therein. The more behavioral approach that motivates $\kappa$-ignorance is, in fact, closely related to the robust control approach. In both cases, the solution to the control problem takes the form of a control band policy.
	However, the robust control of approach of \citet{chakraborty2023optimal} gives rise to a nonlinear Bellman equation, which poses significant challenges for analytically deriving comparative statics results. In contrast, our model admits analytical results by reformulating the classical singular control problem under $\kappa$-ignorance into a Dynkin game, as detailed in Section~\ref{sec:affine}. To the best of our knowledge, this is the first analytical exploration of comparative statics for singular control under ambiguity.
	
	It is important to recognize that while our model is motivated by ambiguity in cash management, it can be easily adapted to models of singular control for other applications. Essentially, our model allows the inventory process to take on various diffusion types, as outlined in Section~\ref{sec:affine}. For instance, while we focus on cash management with a simple arithmetic Brownian motion diffusion, one could apply the same concept to managing, for example, a firm's stock of outstanding shares, where the market maker continuously adjusts shares to maintain a desired price trajectory. This type of inventory process exhibits mean-reversion characteristics (cf. \citet{Cadenillas2010}). Furthermore, our model can extend to non-monetary domains such as dam reservoir control, as demonstrated by \citet{Jiang2022-qs}, where reservoir levels exhibit seasonal mean reversion. Here, ambiguity can be addressed by incorporating seasonal pattern uncertainty, potentially influenced by climate change. Thus, our model represents a general contribution to singular control with maxmin preference ambiguity.
	
	The structure of this paper is as follows: In Section \ref{sec:model} we construct a general formulation for singular control of the Brownian cash reserve under ambiguity. We provide a verification theorem for the optimal control band policy and the existence of the ambiguity trigger in Section \ref{sec:verification}. In Section~\ref{sec:affine} we provide a simplification of the verification theorem and the associated Dynkin game for the case where the present value of the (uncontrolled) expected holding costs is affine in the current value of the cash holdings. This includes, e.g., the case where the uncontrolled cash process follows an arithmetic Brownian motion, or a mean-reverting Ornstein-Uhlenbeck process. A theoretical comparative static analysis for the arithmetic Brownian motion case is given in Section \ref{sec:ABM}.
	
	\section{Simple Cash-Management Model with Drift Ambiguity}\label{sec:model}
	Let $E\subseteq\sr$ be a connected state space endowed with the Euclidean topology and such that $0\in E$. Given $(\Omega,\cF,\Pp)$ a complete probability space. 
	On $(\Omega,\cF,\Pp)$, we assume that $\alpha:E\to\sr$ and $\sigma:E\to\sr$ are continuously differentiable functions such that for all $x\in E$
	\begin{align}
		&\sigma'(x)\text{ is locally Lipschitz continuous} \label{con: loc lips}\\
		&|\alpha(x)| +|\sigma(x)|\leq C(1+|x|) \label{eq: suff con a}
	\end{align}
	for some $C\in \sr$. Then a time-homogeneous diffusion, $X\triangleq\pro{X}$, taking values in $E$, is the unique strong solution to the stochastic differential equation (SDE),
	\begin{equation}
		\rd X_t = \alpha(X_t) \rd t+\sigma(X_t)\rd B_t,\quad \Pp(X_0 = x)=1,
	\end{equation}
	where $B\triangleq\pro{B}$ is a standard Brownian motion. Dynamic revelation of information is modeled by the natural filtration  $\bF=\pro{\cF}$ generated by $X$. We assume that the end points of $E$ are $\Pp$-a.s. unattainable, given $\Pp(X_0 = x)=1$. For brevity of notation we writes $\Pp_x(\cdot)\triangleq \Pp(\cdot|X_0=x)$, associated with an expectation operator $\Ep_x$.
	
	A \emph{control policy} is a pair of processes $(L,U)$, where $L$ and $U$ are adapted, non-decreasing, and non-negative. These processes are associated with increases and decreases, respectively, of $X$ at times at which control is exerted. With the policy $(L,U)$ we associate the \emph{controlled process} $X^{L,U}$ and we say that a control policy $(L,U)$ is \emph{feasible} if for all $x\in E$, there exists a unique $X^{L,U}$ that strongly solves 
	\begin{align}
		\rd X^{L,U}_t &= \alpha(X^{L,U}_t) \rd t+\sigma(X^{L,U}_t)\rd B_t +\rd L_t - \rd U_t, \;\; X_0 = x,\;\; \Pp_x-\text{a.s.},
	\end{align}
	and if there exist $A>0$ and $B<0$, such that
	\begin{align}\label{eq:bounded prob of the controlled X}
		\Pp_x\left(\sup_{t\geq 0}X^{L,U}_t<A, \inf_{t\geq 0}X^{L,U}_t>B\right) &= 1
	\end{align}
	The set of feasible control policies is denoted by $\cD$, while we denote by $X^0$, the uncontrolled process; that is, $X^0\triangleq X^{0,0}$.
	
	The instantaneous holding costs are given by an almost everywhere differentiable function $c:\sr\to\sr_+$. For simplicity we will assume that 
	\begin{equation}
		c(x) = 
		\begin{cases}
			\ch  |x| &\text{if $x\geq 0$}\\
			\cc |x| &\text{if $x< 0$.}
		\end{cases}
	\end{equation}
	for some $\ch  ,\cc >0$. The instantaneous and proportional costs of lower and upper control are denoted by $\ell>0$ and $u>0$, respectively. Our results can easily be extended to more general convex holding costs with $c(0)=0$, albeit at the cost of more cumbersome notation. In a cash management setting, one could think of 0 as the \emph{target level} of cash. When $x>0$, the firm has excess cash while if $x<0$ the firm needs to access cash on the markets. When the cash reserves get too low the firm may need to issue new equity, which incurs costs $\ell$, whereas when $x$ gets too large, the firm may wish to pay out dividends, which incurs a cost $u$.
	
	The DM discounts costs at the constant rate $\rho>0$.
	We, furthermore, assume that
	\begin{equation*}
		\Ep_x\left[\int_0^\infty e^{-\rho t}|X^0_t|\rd t\right]<\infty,\;\;\;x\in E.
	\end{equation*}
	A typical process that satisfies all the assumptions made so far is the arithmetic Brownian motion (ABM), defined on the state space $E=\sr$, being the strong solution of the SDE
	\begin{equation}\label{eq:ABM}
		\dt X^0_t = \alpha \dt t+\sigma \dt B_t,
	\end{equation}
	with constant drift $\alpha\in\sr$ and standard deviation $\sigma>0$. For this specification the uncontrolled cash process is
	\begin{equation*}
		X^0_t = x+\alpha t+\sigma B_t,
	\end{equation*}
	whereas for any feasible control policy $(L,U)\in\cD$, the controlled cash process satisfies
	\begin{equation*}
		X^{L,U,\theta}_t = x+\alpha t+\sigma B_t+L_t-U_t.
	\end{equation*}
	Another process that can be used is the mean-reverting Ornstein-Uhlenbeck (OU) process
	\begin{equation*}
		\dt X^0_t = -\beta X^0_t\dt t+\sigma \dt B_t,
	\end{equation*}
	where $\beta>0$ is the speed of mean-reversion. In this case
	\begin{equation*}
		X^0_t = xe^{-\beta t}+\int_0^te^{-\beta(t-s)}\dt B_s.
	\end{equation*}
	
	It is assumed that the DM is \emph{ambiguous} about the measure $\Pp_x$ and, consequently, considers a set of priors $\mathscr{P}^{\Theta}$. Each of these priors is constructed from the reference measure $\Pp_x$ by means of a \emph{density generator} $\theta\in\Theta$. A process $\theta=\pro{\theta}$ is a density generator if the process $\pro{M^{\theta}}$, with
	\begin{equation}\label{eq:dengen}
		\frac{\rd M_t^{\theta}}{M_t^{\theta}} = -\theta_t\rd B_t,\quad M_0^{\theta}=1,
	\end{equation}
	is a $\Pp_x$--martingale. Such a process $\theta$ generates a new measure $\Pp_x^{\theta}$ on $(\Omega,\cF^{B})$ via the Radon--Nikodym derivative $\rd\Pp_x^{\theta}/\rd\Pp_x\big|_{\cF^{B}_T}=M^{\theta}_T$ for any $T>0$. Here, $\cF^{B}\triangleq\cF^{B}_{\infty}$, where $\bF^{B}\triangleq\pro{\cF^{B}}$ is the (uncompleted) filtration generated by $B$. Indeed, if $\theta\in\Theta$, then it follows from Girsanov's theorem (see, Corollary 5.2 in Chapter 3.5 of \citealp{karatzas1991brownian}) that under the measure $\Pp_x^\theta$ the process $B^\theta\triangleq\pro{B^{\theta}}$, defined by
	\begin{equation}\label{def:B theta}
		B^{\theta}_t \triangleq B_t+\int_0^t\theta_s\rd s,
	\end{equation}
	is a Brownian motion on $(\Omega, \cF^{B}, \bF^{B}, \Pp^{\theta}_x)$ and that, under $\Pp_x^{\theta}$, the process $X^{L,U,\theta}$ is the unique strong solution to the SDE
	\begin{align*}
		\rd X^{L,U,\theta}_t &=\alpha(X^{L,U,\theta}_t)\rd t+\sigma(X^{L,U,\theta}_t) \rd B^{\theta}_t +\rd L_t - \rd U_t\\
		&= \Big(\alpha(X^{L,U,\theta}_t)-\sigma(X^{L,U,\theta}_t)\theta_t\Big)\rd t+\sigma(X^{L,U,\theta}_t) \rd B_t +\rd L_t - \rd U_t, \quad\Pp_x-\text{a.s}.
	\end{align*}
	In the remainder we restrict attention to so-called \emph{$\kappa$-ignorance}, i.e., we only use density generators $\theta$ for which $\theta_t\in[-\kappa,+\kappa]$ for all $t\geq 0$ and some $\kappa\geq 0$. Note that $\mathscr{P}^{\Theta}=\{\Pp_x\}$ if $\kappa=0$.  
	
	To model \emph{ambiguity aversion}, it is assumed that the DM uses maxmin utility \emph{\`{a} la} \cite{GiSch89}. That is, the \emph{worst-case cost function} associated with the feasible policy $(L,U)\in\cD$ is given by $J^{L,U}:E\to\sr$, where
	\begin{equation}
		J^{L,U}(x) \triangleq \sup_{\theta\in\Theta}\Ep^\theta_x\left[\int_0^{\infty}e^{-\rho t}\left(c(X^{L,U,\theta}_t)\rd t+\ell \rd L_t+u \rd U_t\right)\right].
	\end{equation}
	The DM's objective is to find the feasible policy that minimizes the worst-case expected costs over the set of priors $\cP^\Theta$. The firm's \emph{minimal cost function} is
	\begin{equation}\label{eq:Value}
		V(x) \triangleq \inf_{(L,U)\in\cD}J^{L,U}(x).
	\end{equation}
	From \citet[Theorem~2.1]{ChEp02} it follows that there exists an \emph{upper-rim generator} $\theta^\ast\in\Theta$ so that
	\begin{equation}\label{eq:J_upper_rim}
		J^{L,U}(x) = \Ep^{\theta^\ast}_x\left[\int_0^{\infty}e^{-\rho t}\left\{c\big(X^{L,U,\theta^\ast}_t\big)\rd t+\ell \rd L_t+u \rd U_t\right\}\right]. 
	\end{equation}
	Furthermore, from \citet[Section~3.3]{ChEp02} it follows that under $\kappa$-ignorance it holds that $\theta^\ast_t\in\{-\kappa,\kappa\}$ for all $t\geq 0$.
	
	Finally, in many cases the optimal policy consists of exerting control only when the process $X$ exits an interval $(\xL,\xU)$. Therefore, with each pair $(\xL,\xU)\in E\times E$, $\xL<\xU$, we associate the \emph{control band policy} $(L,U)\in\cD$ for which $\xL$ is an (upward) reflecting barrier for $L$ and $\xU$ is a (downward) reflecting barrier for $U$.  
	For such policies it holds that
	\begin{enumerate}
		\item $X^{L,U,\theta^\ast}_t\in[\xL,\xU]$, $\Pp_x$-a.s. for all $t\geq 0$, and
		
		\item $\int_0^\infty 1_{(\xL,\xU)}(X^{L,U,\theta^\ast}_t)\rd(L_t+U_t)=0$, $\Pp_x$-a.s.
	\end{enumerate}
	Following \citet{tanaka1979stochastic}, our assumptions on $X$ are sufficient to guarantee the existence of control band policies.

	\section{A General Verification Theorem}\label{sec:verification}
	
	Let $\cL$ denote the characteristic operator on $C^2(E)$ of the killed  process $\big(e^{-\rho t}X_t\big)_{t\geq 0}$ under $\Pp_x$, i.e.
	\begin{equation}
		\cL\varphi(x) = \frac{1}{2}\sigma^2(x)\varphi''(x)+\alpha(x)\varphi'(x)-\rho\varphi(x).
	\end{equation}
	On $C^1(E)$ we also define the density generator,
	\begin{equation*}
		\theta^\varphi\triangleq\theta^\varphi(x)\triangleq
		\begin{cases}
			-\kappa &\text{if $\varphi'(x)\geq 0$}\\
			+\kappa &\text{if $\varphi'(x)<0$.}
		\end{cases}
	\end{equation*}
	We get the following verification theorem.
	\begin{theorem}\label{th:verification}
		Suppose there exists a pair $(\xL,\xU)\in E\times E$, $\xL<0<\xU$, and a non-negative, convex, and $C^2$-function $\varphi$ on $(\xL,\xU)$ such that
		\begin{enumerate}
			\item $\theta^\varphi(x)\sigma(x)\varphi'(x)-\cL\varphi(x)=c(x)$ on $(\xL,\xU)$,
			
			\item $\varphi'(\xL+)=-\ell$, $\varphi'(\xU-)=u$,
			
			\item $\varphi''(\xL+)=\varphi''(\xU-)=0$, 
			
			
			
			\item $\cc (\xL-x)\geq\ell\big[\rho(\xL-x)-\big(\alpha(\xL)-\alpha(x)\big)-\kappa\big(\sigma(\xL)-\sigma(x)\big)\big]$, for all $x<\xL$, 
			
			\item $\ch  (x-\xU)\geq u\big[\rho(x-\xU)-\big(\alpha(x)-\alpha(\xU)\big)+\kappa\big(\sigma(x)-\sigma(\xU)\big)\big]$, for all $x>\xU$, and
			
			\item $\lim_{T\to\infty}e^{-\rho T}\Ep_x^{\theta^\varphi}\left[\varphi\big(X_T^{L,U}\big)\right]=0$, for all $(L,U)\in\cD$.
		\end{enumerate}
		Then the optimal policy $(L^\ast,U^\ast)$ is the control band policy associated with $(\xL,\xU)$ and the minimal cost function is 
		\begin{equation*}
			V(x) =  
			\begin{cases}
				\ell|\xL-x|+\varphi(\xL+) &\text{if $x\leq\xL$}\\
				\varphi(x) &\text{if $\xL<x<\xU$}\\
				u|x-\xU|+\varphi(\xU-) &\text{if $x\geq\xU$}
			\end{cases}.
		\end{equation*}
	\end{theorem}
	
	\begin{remark}
		The optimal control policy $(L^\ast,U^\ast)$ is such that no control takes place as long as the cash stock is between $\xL$ and $\xU$. This implies that the processes $L^\ast$ and $U^\ast$ remain constant. Whenever the lower bound $\xL$ is hit, the processes $L^\ast$ increases to keep the cash stock at $\xL$. Similarly, whenever the lower bound $\xU$ is hit, the processes $U^\ast$ increases to keep the cash stock at $\xU$.
	\end{remark}
	
	\begin{remark}
		Conditions 4 and 5 guarantee existence of a feasible policy under ambiguity. For the case of an uncontrolled ABM,
		\begin{equation*}
			\dt X^0 = \alpha \dt t+\sigma \dt B_t,
		\end{equation*}
		these conditions reduce to 
		\begin{equation*}
			\cc \geq \rho\ell,\quad\text{and}\quad\ch  \geq\rho u. 
		\end{equation*}
		That is, the discounted perpetual holding costs of positive (negative) cash balances should exceed the control costs of reducing (increasing) the cash balance. As another example, for the case of an uncontrolled mean-reverting OU process,
		\begin{equation*}
			\dt X^0 = \beta(\tilde{x}-X^0_t) \dt t+\sigma \dt B_t,
		\end{equation*}
		where $\tilde{x}$ is the long-run mean and $\beta$ is the speed of mean reversion, conditions 4 and 5 reduce to 
		\begin{equation*}
			\cc \geq (\rho+\beta)\ell,\quad\text{and}\quad\ch  \geq(\rho+\beta)u. 
		\end{equation*}
	\end{remark}
	
	\begin{proof}[Proof of Theorem~\ref{th:verification}]
		Let $\varphi$ and $\xL<0<\xU$ satisfy conditions 1--6. Extend $\varphi$ to $E$, in a twice-continuously differentiable way, as follows:
		\begin{equation}\label{eq:extended varphi}
			\varphi(x) = 
			\begin{cases}
				\ell|\xL-x|+\varphi(\xL+) &\text{if $x\leq\xL$}\\
				\varphi(x) &\text{if $\xL<x<\xU$}\\
				u|x-\xU|+\varphi(\xU-) &\text{if $x\geq\xU$}
			\end{cases}.
		\end{equation}
		Let $(L^\ast,U^\ast)$ be the control band policy associated with $(\xL,\xU)$. The proof proceeds in two steps. First we prove that $J^{L^\ast,U^\ast}=\varphi$. Then we show that for any other feasible policy $(L,U)$ it holds that $J^{L,U}\geq J^{L^\ast,U^\ast}$, so that $J^\ast=J^{L^\ast,U^\ast}$. Note that 
		\begin{equation}\label{eq:worst-case1}
			\theta^\varphi(x) = {\arg\min}_{\theta\in[-\kappa,+\kappa]}\big(\theta\sigma(x)\text{sign}(\varphi'(x))\big),
		\end{equation}
		so that the worst-case prior is generated by
		\begin{equation}\label{eq:worst-case2}
			\theta^\ast_t(\omega) = \theta^\varphi\big(X_t(\omega)\big).
		\end{equation}
		
		\noindent \textbf{1.}  Fix $T>0$, $x\in E$, $\theta\in\Theta$ and set $\theta^\varphi_t \triangleq \theta^\varphi(X^{L^\ast,U^\ast}_t)$. 
		From It\={o}'s lemma it then follows that
		\begin{align*}
			&\Ep^{\theta}_x\left[e^{-\rho T}\varphi\big(X^{L^\ast,U^\ast,\theta}_T\big)\right] 
			\\&= \varphi(x)+\Ep^{\theta}_x\bigg[\int_0^Te^{-\rho s}\varphi'\big(X^{L^\ast,U^\ast,\theta}_s\big)\rd\left(L^\ast_s+U^\ast_s\right) \\&\quad\quad\quad\quad\quad\quad +\int_0^T e^{-\rho s} \bigg\{\cL\varphi\big(X^{L^\ast,U^\ast,\theta}_s\big)-\theta_s\sigma\big(X^{L^\ast,U^\ast,\theta}_s\big) \varphi'\big(X^{L^\ast,U^\ast,\theta}_s\big)\bigg\}\rd s\Bigg]\\
			&\leq \varphi(x)+\Ep^{\theta}_x\Bigg[\int_0^Te^{-\rho s}\varphi'\big(X^{L^\ast,U^\ast,\theta}_s\big)\rd\left(L^\ast_s+U^\ast_s\right)\\&\quad\quad\quad\quad\quad\quad\quad +\int_0^T e^{-\rho s} \big\{\cL\varphi\big(X^{L^\ast,U^\ast,\theta}_s\big)-\theta^\varphi_s\sigma\big(X^{L^\ast,U^\ast,\theta}_s\big) \varphi'\big(X^{L^\ast,U^\ast,\theta}_s\big)\big\}\rd s\Bigg]\\
			&=\varphi(x)-\Ep_x^\theta\Bigg[\int_0^Te^{-\rho s}c\left(X_s^{L^\ast,U^\ast,\theta}\right)\dt s+\int_0^Te^{-\rho s}\left(u\dt U^\ast_s+\ell\dt L^\ast_s\right)\Bigg],
		\end{align*}
		where the inequality follows from ~\eqref{eq:worst-case1} and ~\eqref{eq:worst-case2}, and the final equality follows from conditions~1 and~2.
		
		Sending $T\to\infty$ and exploiting the non-negativity of $\varphi$, $c$, $u$, and $\ell$, we find that
		\begin{align*}
			\varphi(x) &\geq \Ep_x^\theta\Bigg[\int_0^{\infty}e^{-\rho s}c\left(X_s^{L^\ast,U^\ast,\theta}\right)\dt s+\int_0^{\infty}e^{-\rho s}\left(u\dt U^\ast_s+\ell\dt L^\ast_s\right)\Bigg].
		\end{align*}
		Since $\theta\in\Theta$ was chosen arbitrarily, this implies that
		\begin{align*}
			\varphi(x)
			&\geq \inf_{(L,U)\in\cD}\sup_{\theta\in\Theta}\Ep_x^\theta\Bigg[\int_0^{\infty}e^{-\rho s}c\left(X_s^{L,U,\theta}\right)\dt s+\int_0^{\infty}e^{-\rho s}\left(u\dt U_s+\ell\dt L_s\right)\Bigg].
		\end{align*}
		
		\noindent\textbf{2.} Next, note that Conditions~4 and~5 ensure that
		\begin{equation}\label{eq: var ineq}
			\theta^\varphi(\xL)\sigma(x)\varphi'(x)-\cL\varphi(x) \leq c(x),\quad\text{on $E$.}
		\end{equation}
		On $(\xL,\xU)$ this holds by construction. To see that it holds for $x\leq\xL$, note that condition~4 and \eqref{eq:extended varphi} implies that
		\begin{align*}
			c(x)=-\cc x &\geq \rho\ell(\xL-x) -\cc \xL - \ell(\alpha(\xL)-\alpha(x)-\kappa(\sigma(\xL)-\sigma(x))\\
			&=\rho(\varphi(x)-\varphi(\xL)) -\cc \xL +\varphi'(\xL)(\alpha(\xL)-\kappa\sigma(\xL)) - \varphi'(x)(\alpha(x)-\kappa\sigma(x)) \\
			&= \theta^\varphi(\xL)\sigma(x)\varphi'(x)-\cL\varphi(x)
		\end{align*}
		The equalities hold since $\varphi'(x)=-\ell,\varphi''(x)=0$ for any $x\leq\xL$ and that $\theta^{\varphi}(\xL)=\kappa$. Similarly, Condition~5 ensures that the results holds for $x\geq\xU$. Then, from convexity it follows that
		\begin{equation}\label{eq:deriv_bounds}
			-\ell\leq\varphi'(x)\leq u,\quad\text{on $E$.} 
		\end{equation}
		
		\noindent\textbf{3.} Let $(\bar{L},\bar{U})$ be a feasible control policy. Fix $T>0$. An application of It\={o}'s lemma now gives that
		\begin{align*}
			&\Ep_x^{\theta^\varphi}\Bigg[\int_0^T e^{-\rho s}\Big\{c(X^{\bar{L},\bar{U},\theta^\varphi}_s)\rd s+u\rd\bar{U}_s+\ell\rd\bar{L}_s\Big\}\Bigg]\\&
			\underset{\eqref{eq:deriv_bounds}}{\geq} \Ep_x^{\theta^\varphi}\Bigg[\int_0^T e^{-\rho s}\Big\{\theta^\varphi\big(X_s^{\bar{L},\bar{U},\theta^\varphi}\big)\sigma\big(X_s^{\bar{L},\bar{U},\theta^\varphi}\big) \varphi'\big(X_s^{\bar{L},\bar{U},\theta^\varphi}\big)-\cL\varphi\big(X_s^{\bar{L},\bar{U},\theta^\varphi}\big)\Big\}\rd t\\
			&\quad\quad\quad\quad\quad\quad-\int_0^T e^{-\rho s}\varphi'\big(X_s^{\bar{L},\bar{U},\theta^\varphi}\big)\big(\rd\bar{L}_s - \rd\bar{U}_s\big)\Bigg]\\
			&=\varphi(x)-\Ep_x^{\theta^\varphi}\left[e^{-\rho T}\varphi\big(X_s^{\bar{L},\bar{U},\theta^\varphi}\big)\right].
		\end{align*}
		Therefore,
		\begin{align*}
			\varphi(x) &\leq \Ep_x^{\theta^\varphi}\Bigg[\int_0^T e^{-\rho s}\Big\{c(X^{\bar{L},\bar{U},\theta^\varphi}_s)\rd s+u\rd\bar{U}_s+\ell\rd\bar{L}_s\Big\}e^{-\rho T}\varphi\big(X_s^{\bar{L},\bar{U},\theta^\varphi}\big)\Bigg].
		\end{align*}
		Sending $T\to\infty$ and by exploiting Condition~6, the monotone convergence theorem gives
		\begin{align*}
			\varphi(x) &\leq \Ep_x^{\theta^\varphi}\Bigg[\int_0^{\infty} e^{-\rho s}\Big\{c(X^{\bar{L},\bar{U},\theta^\varphi}_s)\rd s+u\rd\bar{U}_s+\ell\rd\bar{L}_s\Big\}\Bigg]
		\end{align*}
		By arbitrariness of $(\bar{L},\bar{U})$, it then follows that
		\begin{align*}
			\varphi(x) &\leq \sup_{\theta\in\Theta}\inf_{(L,U)\in\cD}\Ep_x^{\theta}\Bigg[\int_0^{\infty} e^{-\rho s}\Big\{c(X^{L,U,\theta}_s)\rd s+u\rd U_s+\ell\rd L_s\Big\}\Bigg]\\
			&=\sup_{\theta\in\Theta}\Ep_x^{\theta}\Bigg[\int_0^{\infty} e^{-\rho s}\Big\{c(X^{L^*,U^*,\theta}_s)\rd s+u\rd U^*_s+\ell\rd L^*_s\Big\}\Bigg].
		\end{align*}
		
		\noindent\textbf{4.}
		Combining the results from Steps~1 and~3 gives that 
		\begin{align*}
			\varphi(x)&=\inf_{L,U\cD}\sup_{\theta\in\Theta}\Ep^{\theta}_x\Bigg[ \int_0^{\infty} e^{-\rho s}\Big\{c(X^{L,U,\theta}_s)\rd s+u\rd U_s+\ell\rd L_s\Big\} \Big]\\
			&= \sup_{\theta\in\Theta}\inf_{L,U\in\cD}\Ep^{\theta}_x\Bigg[ \int_0^{\infty} e^{-\rho s}\Bigg\{c(X^{L,U,\theta}_s)\rd s+u\rd U_s+\ell\rd L_s\Big\} \Bigg]
		\end{align*}
		and that $(\theta^{\varphi},(L^{\ast},U^{\ast}))$ realise a saddle-point. 
	\end{proof}
	
	\section{Affine Perpetual Holding Costs}\label{sec:affine}
	Under some additional assumptions, it is often possible to write down sufficient conditions that are easier to check. In order to pursue this program, we first derive an expression for the perpetual holding costs of the \emph{uncontrolled} process. First we let $\hat{\varphi}_{\pm\kappa}$ and $\check{\varphi}_{\pm\kappa}$ denote the increasing and decreasing  \emph{fundamental} solutions to the ordinary differential equation (ODE)
	\begin{equation*}
		\cL\varphi(x)-\theta^{\pm\kappa}(x)\sigma(x)\varphi'(x)=0,\quad\text{on $E$,}
	\end{equation*}
	respectively. Here $\theta^{\pm\kappa}$ is the density generator $\theta^{\pm\kappa}(x)=\pm\kappa$, all $x\in E$. The measure generated by $\theta^{\pm\kappa}$ is denoted by $\Pp^{\pm\kappa}_x$. We normalize $\hat{\varphi}_{\pm\kappa}(0)=\check{\varphi}_{\pm\kappa}(0)=1$, and denote
	\begin{equation*}
		f_{\pm\kappa}(x) \triangleq \Ep_x^{\theta^{\pm\kappa}}\left[\int_0^\infty e^{-\rho t}X^0_t\dt t\right],
	\end{equation*}
	where we assume that $f_{\pm\kappa}$ is affine in $x$. We summarize our assumptions on $X^0$ for future reference.
	
	\begin{assumption}\label{ass:X}
		The process $X^0$ is such that
		\begin{enumerate}
			\item the present value of its expected evolution is affine in its current state, i.e.
			\begin{equation}\label{eq:f_affine}
				f_{\pm\kappa}(x)=\Ep_x^{\theta^{\pm\kappa}}\left[\int_0^\infty e^{-\rho t}X^0_t\dt t\right] = ax+b,\quad(a\neq 0),
			\end{equation}
			
			\item the increasing and decreasing solutions, $\hat{\varphi}$ and $\check{\varphi}$, to $\cL\varphi(x)-\theta^{\pm\kappa}(x)\sigma(x)\varphi'(x)=0$ are convex (see \citealp{Alvarez2003} for sufficient conditions) and such that
			\begin{equation}\label{eq:fundamental_solution}
				\hat{\varphi}(0) = \check{\varphi}(0)=1,
			\end{equation}
			and
			
			\item the holding costs $\cc $ and $\ch  $ are such that
			\begin{equation}\label{eq:cost_condition}
				\ell\leq \cc f'_{-\kappa}(x), \quad\text{and}\quad
				u\leq \ch  f'_{+\kappa}(x).
			\end{equation}
		\end{enumerate}
	\end{assumption}
	
	It can be verified that both ABM and OU process satisfy this assumption. For example, if $X^0$ follows an ABM, then
	\begin{equation*}
		f_{\pm\kappa}(x) = \frac{x}{\rho}+\frac{\alpha\pm\kappa\sigma}{\rho^2}.
	\end{equation*}
	Moreover, 
	\begin{equation*}
		\hat{\varphi}_{\pm\kappa}(x) = e^{\beta_{\pm\kappa}x}, \quad\text{and}\quad \check{\varphi}_{\pm\kappa}(x) = e^{\gamma_{\pm\kappa}x},
	\end{equation*}
	where $\beta_{\pm\kappa}>0$ and $\gamma_{\pm\kappa}<0$ are the roots of the quadratic equation
	\begin{equation*}
		\cQ_{\pm\kappa}(\chi)\equiv\frac{1}{2}\sigma^2\chi^2 +(\alpha\pm\kappa\sigma)\chi-\rho = 0.
	\end{equation*}
	Condition~\eqref{eq:cost_condition} now reduces to
	\begin{equation*}
		\ell\leq \cc /\rho, \quad\text{and}\quad
		u\leq \ch  /\rho.
	\end{equation*}
	
	If, on the other hand, $X^0$ follows the mean-reverting process
	\begin{equation*}
		\dt X^0_t = -\beta X^0_t\dt t+\sigma\dt B_t,\quad(\eta>0),
	\end{equation*}
	under $\Pp_x$, then under $\Pp^{\pm\kappa}_x$ it holds that,
	\begin{equation*}
		\dt X^0_t = (-\beta X^0_t\pm\kappa\sigma)\dt t+\sigma\dt B^{\pm\kappa}_t,
	\end{equation*}
	where $B^{\pm\kappa}$ is a $\Pp^{\pm\kappa}_x$-Brownian motion. This process can be seen as an Ornstein-Uhlenbeck (OU) process with long-run mean $\tilde{x}_{\pm\kappa}$, i.e.
	\begin{align*}
		\dt X^0_t &= \beta(\tilde{x}_{\pm\kappa}-X^0_t)\dt t+\sigma\dt B^{\pm\kappa}_t,\quad \text{where   }\tilde{x}_{\pm\kappa}=\pm\kappa\sigma/\beta.
	\end{align*}
	Therefore,
	\begin{equation*}
		f_{\pm\kappa}(x) = \frac{x-\tilde{x}_{\pm\kappa}}{\rho+\beta}+\frac{\tilde{x}_{\pm\kappa}}{\rho}.
	\end{equation*}
	
	Here, the fundamental solutions of which $X^0$ follows the OU process are
	\begin{align*}
		\hat{\varphi}_{\pm\kappa}(x) = e^{\frac{\beta(x-\tilde x_{\pm\kappa})^2}{2\sigma^2}}D_{-\frac{\rho}{\beta}}\Big(\frac{x-\tilde x_{\pm\kappa}}{\sigma}\sqrt{2\beta}\Big), \quad\text{and}\quad \check{\varphi}_{\pm\kappa}(x) = e^{\frac{\beta(x-\tilde x_{\pm\kappa})^2}{2\sigma^2}}D_{-\frac{\rho}{\beta}}\Big(-\frac{x-\tilde x_{\pm\kappa}}{\sigma}\sqrt{2\beta}\Big),
	\end{align*}
	where $D_{z}$ is the parabolic cylinder function with index $z$ (see, for example, \citet{Jeanblanc2009-re}, chapter 5).
	
	Condition~\eqref{eq:cost_condition} now reduces to
	\begin{equation*}
		\ell\leq \cc /(\rho+\beta), \quad\text{and}\quad
		u\leq \ch  /(\rho+\beta).
	\end{equation*}
	
	The perpetual holding costs of the uncontrolled process can be found using the Feynman-Kac formula in the standard way:
	\begin{align*}
		R_{\pm\kappa}(x) &\triangleq \Ep_x^{\theta^{\pm\kappa}}\left[\int_0^\infty e^{-\rho t}c(X^0_t)\dt t\right]
		=
		\begin{cases}
			-\cc f_{\pm\kappa}(x) +\hat{E}_{\pm\kappa}\hat{\varphi}_{\pm\kappa}(x) &\text{if $x<0$}\\
			+\ch  f_{\pm\kappa}(x) +\check{E}_{\pm\kappa}\check{\varphi}_{\pm\kappa}(x) &\text{if $x\geq 0$.}
		\end{cases}
	\end{align*}
	Here, $\hat{E}_{\pm\kappa}$ and $\check{E}_{\pm\kappa}$ are constants that are determined by ``value-matching'' and ``smooth-pasting'' conditions at 0, i.e.,
	\begin{align*}
		R_{\pm\kappa}(0-)=R_{\pm\kappa}(0+),\;\;\text{and}\;\; R'_{\pm\kappa}(0-)=R'_{\pm\kappa}(0+),
	\end{align*}
	respectively. This gives 
	\begin{align*}
		&\hat{E}_{\pm\kappa} = (\ch  +\cc )\frac{f'_{\pm\kappa}(0)-f_{\pm\kappa}(0)\check{\varphi}'_{\pm\kappa}(0)}{\hat{\varphi}'_{\pm\kappa}(0)-\check{\varphi}'_{\pm\kappa}(0)},\quad\text{and}\\& \check{E}_{\pm\kappa} = (\ch  +\cc )\frac{f'_{\pm\kappa}(0)-f_{\pm\kappa}(0)\hat{\varphi}'_{\pm\kappa}(0)}{\hat{\varphi}'_{\pm\kappa}(0)-\check{\varphi}'_{\pm\kappa}(0)},
	\end{align*}
	so that
	\begin{align*}
		&R_{\pm\kappa}(x)=
		\begin{cases}
			-\cc f_{\pm\kappa}(x) + (\ch  +\cc )\frac{f'_{\pm\kappa}(0)-f_{\pm\kappa}(0)\check{\varphi}'_{\pm\kappa}(0)}{\hat{\varphi}'_{\pm\kappa}(0)-\check{\varphi}'_{\pm\kappa}(0)}\hat{\varphi}_{\pm\kappa}(x) &\text{if $x<0$}\\
			+\ch  f_{\pm\kappa}(x) +(\ch  +\cc )\frac{f'_{\pm\kappa}(0)-f_{\pm\kappa}(0)\hat{\varphi}'_{\pm\kappa}(0)}{\hat{\varphi}'_{\pm\kappa}(0)-\check{\varphi}'_{\pm\kappa}(0)}\check{\varphi}_{\pm\kappa}(x) &\text{if $x\geq 0$.}
		\end{cases}
	\end{align*}
	
	Without ambiguity ($\kappa=0$), in order to construct the function $\varphi$ of Theorem~\ref{th:verification}, one would now find constants $A$ and $B$, and control barriers $\xL$ and $\xU$ such that the following value-matching and smooth-pasting conditions hold:
	\begin{align*}
		R'_0(\xL)+A\hat{\varphi}'_0(\xL)+B\check{\varphi}'_0(\xL) &= -\ell\\
		R'_0(\xU)+A\hat{\varphi}'_0(\xU)+B\check{\varphi}'_0(\xU) &= u\\
		R''_0(\xL)+A\hat{\varphi}''_0(\xL)+B\check{\varphi}''_0(\xL) &= 0\\
		R''_0(\xU)+A\hat{\varphi}''_0(\xU)+B\check{\varphi}''_0(\xU) &= 0.
	\end{align*}
	One then proceeds by showing that the resulting function,
	\begin{equation*}
		\varphi(x) = R_0(x)+A\varphi_0(x)+B\varphi_0(x),\quad\text{on $(\xL,\xU)$,}
	\end{equation*}
	and the constants $\xL<0<\xU$ satisfy the conditions of the verification Theorem~\ref{th:verification}.
	
	Under ambiguity ($\kappa>0$) matters are a bit more complicated. Intuitively speaking, the main issue is that the ``worst--case drift'' is different at $\xL$ and $\xU$. In particular, at the lower control bound the worst case drift is $\alpha(\xL)-\kappa\sigma(\xL)$, because the worst that can happen is that the cash hoard depletes even more and, thus, increases the control costs. Similarly, at the upper control bound the worst case drift is $\alpha(\xU)+\kappa\sigma(\xU)$, because the worst that can happen is that the cash hoard increases even more and, thus, increases the control costs. 
	
	So, at $\xL$ and $\xU$ we need to work with functions $R$, $\hat{\varphi}$, and $\check{\varphi}$ under $-\kappa$ and $+\kappa$, respectively. That is, we will look for constants $A$, $B$, $C$, and $D$, as well as control bounds $\xL$ and $\xU$ such that the following value-matching and smooth-pasting conditions hold:
	\begin{align}
		R'_{-\kappa}(\xL)+A\hat{\varphi}'_{-\kappa}(\xL)+B\check{\varphi}'_{-\kappa}(\xL) &= -\ell \label{eq:VM_xL}\\
		R''_{-\kappa}(\xL)+A\hat{\varphi}''_{-\kappa}(\xL)+B\check{\varphi}''_{-\kappa}(\xL) &= 0 \label{eq:SP_xL}\\
		R'_{+\kappa}(\xU)+C\hat{\varphi}'_{+\kappa}(\xU)+D\check{\varphi}'_{+\kappa}(\xU) &= u \label{eq:VM_xU}\\
		R''_{+\kappa}(\xU)+C\hat{\varphi}''_{+\kappa}(\xU)+D\check{\varphi}''_{+\kappa}(\xU) &= 0. \label{eq:SP_xU}
	\end{align}
	
	Now, of course, we have too few equations to determine all the constants. The ``missing'' constraints come from the fact that there is a point $x^\ast$ where the worst-case drift changes. This is the point where the firm's cost function changes from being decreasing to increasing. At this point we also impose a value-matching and smooth-pasting condition, i.e., we find $x^\ast\in(\xL,\xU)$ such that
	\begin{align}
		&R'_{-\kappa}(x^\ast-)+A\hat{\varphi}'_{-\kappa}(x^\ast-)+B\check{\varphi}'_{-\kappa}(x^\ast-) = 0 \label{eq:VM_x*-}\\
		&R'_{+\kappa}(x^\ast+)+C\hat{\varphi}'_{+\kappa}(x^\ast+)+D\check{\varphi}'_{+\kappa}(x^\ast+) = 0 \label{eq:VM_x*+}\\
		&R''_{-\kappa}(x^\ast-)+A\hat{\varphi}''_{-\kappa}(x^\ast-)+B\check{\varphi}''_{-\kappa}(x^\ast-) \nonumber\\
		&\quad =R''_{+\kappa}(x^\ast+)+C\hat{\varphi}''_{+\kappa}(x^\ast+)+D\check{\varphi}''_{+\kappa}(x^\ast+). \label{eq:SP_x*}
	\end{align}
	We show below that if this system of 7 equations in 7 unknowns has a solution, then a function $\varphi$ can be constructed on $(\xL,\xU)$ so that the conditions of verification Theorem~\ref{th:verification} are satisfied. A similar approach has also been used by \cite{Cheng2013-to} to price a straddle option under ambiguity and by \cite{Hellmann2018-xs} to analyse preemptive investment behavior in a duopoly under ambiguity.
	
	\begin{theorem}\label{th:verification2}
		Suppose that the system of equations~\eqref{eq:VM_xL}--\eqref{eq:SP_x*} admits a solution $(A,B,C,D,\xL,\xU,x^\ast)$ with $\xL<x^\ast<\xU$. Then the optimal policy $(L^\ast,U^\ast)$ is the control band policy associated with $(\xL,\xU)$ and the firm's cost function is 
		\begin{equation}\label{J}
			V(x) =  
			\begin{cases}
				\ell(\xL-x)+R_{-\kappa}(\xL+)+A\vH_{-\kappa}(\xL+)+B\vC_{-\kappa}(\xL+) &\text{if}\;\;\;x\leq\xL\\
				R_{-\kappa}(x)+A\vH_{-\kappa}(x)+B\vC_{-\kappa}(x) &\text{if} \;\;\;\xL<x<x^\ast\\
				R_{+\kappa}(x)+C\vH_{+\kappa}(x)+D\vC_{+\kappa}(x) &\text{if} \;\;\;x^\ast\leq x<\xU\\
				u(x-\xU)+R_{+\kappa}(\xU-)+C\vH_{+\kappa}(\xU-)+D\vC_{+\kappa}(\xU-) &\text{if} \;\;\;x\geq\xU.
			\end{cases}
		\end{equation} 
	\end{theorem}
	\begin{proof}[Proof of Theorem \ref{th:verification2}]
		See Appendix~\ref{pf:verification2}.
	\end{proof}
	
	Now we propose a probabilistic representation of the first derivative of the value function, formulated as an optimal stopping game for an uncontrolled process. This approach will be highly beneficial for the comparative statics analysis in the following section. Such a problem is known as a zero-sum Dynkin game for singular control, initially introduced by \citet{Taksar1985-ab}. For further applications in operations research, see also \citet{Guo2008-da, Ferrari2020-oh, Federico2023-pm}.
	\begin{theorem}\label{prop:dynkin}
		Let $V(x)$ be the value function of~\eqref{J} and assume furthermore that $\rho>\sup_{x\in E}|\alpha'(x)| \pm\kappa|\sigma'(x)|$. Then for any $x\in E$, we have
		\begin{align}\label{eq:dynkin}
			V'(x)\triangleq v(x)=\inf_\eta\sup_\tau w(x,\tau,\eta)=\sup_\tau \inf_\eta w(x,\tau,\eta)
		\end{align}
		where $\tau$ and $\eta$ are $\cF_t$-stopping times under $\textup\Pp^{\theta^V}_x$,
		
		\begin{align*}
			w(x,\tau,\eta)&\triangleq\Ep^{\theta^V}_x\Bigg[\int_0^{\tau\wedge\eta}e^{-\int_0^t\hat\rho(\hat X^{\theta^V}_s)\rd s}\zeta(\hat X^{\theta^V}_t)\rd t \nonumber \\ &\quad\quad\quad\quad\quad\quad\quad\quad- 1_{(0,\eta)}(\tau)e^{-\int_0^\tau\hat\rho(\hat X^{\theta^V}_s)\rd s}\ell + 1_{(0,\tau)}(\eta)e^{-\int_0^\eta\hat\rho(\hat X^{\theta^V}_s)\rd s}u\Bigg],\\
			\theta^V(x)&\triangleq\kappa1_{(-\infty,\xS)}(x)-\kappa1_{[\xS,\infty)}(x),\label{def:thetaV}\\
			\zeta(x)&\triangleq\ch  \cdot 1_{(0,\infty)}(x)-\cc \cdot 1_{(-\infty,0)}(x) \;\;\text{and},\\
			\hat\rho(x) &\triangleq \rho - (\alpha'(x) - \theta^V(x)\sigma'(x)).
		\end{align*}
		Here $\hat X^{\theta^V}$ solves
		\begin{align}
			\rd \hat X^{\theta^V}_t &=\left(\alpha(\hat X^{\theta^V}_t) + \sigma'(\hat X^{\theta^V}_t)\sigma(\hat X^{\theta^V}_t) \right)\rd t +\sigma(\hat X^{\theta^V}_t)\rd B^{\theta^V}_t,\nonumber\\
			&= \left(\alpha(\hat X^{\theta^V}_t) + (\sigma'(\hat X^{\theta^V}_t)-\theta^V(\hat X^{\theta^V}_t))\sigma(\hat X^{\theta^V}_t) \right)\rd t +\sigma(\hat X^{\theta^V}_t)\rd B_t,\;\;\;\textup\Pp^{\theta^V}_x\text{-a.s.}
		\end{align}
		where $B^{\theta^V}$ is a Brownian motion in the sense of \eqref{def:B theta} when $\theta=\theta^V$. Moreover, the saddle-point stopping times $(\tau^\ast,\eta^\ast)$ are given by 
		\begin{align}\label{eq:saddle-point}
			\ts &\triangleq \inf\{t>0:\hat X^{\theta^V}_t\leq \xL\}, \;\;\;\text{and}  \;\;\; \es \triangleq \inf\{t>0:\hat X^{\theta^V}_t\geq \xU\},\;\;\textup\Pp^{\theta^V}_x\text{-a.s.}
		\end{align}
		That is,
		\begin{align}
			w(x,\tau,\es)&\leq v(x)\leq w(x,\ts,\eta) \;\;\text{for any}\;\;x\in E
		\end{align}
	\end{theorem} 
	
	\begin{proof}[Proof of Theorem~\ref{prop:dynkin}]
		The idea of what follows is adopted from \citet{Ferrari2020-oh}. One can see from Theorem~\ref{th:verification2} that $V''$ is bounded in $E$, non-decreasing in $E\setminus(x^\ast,\infty)$ and non-increasing in $E\setminus(-\infty,x^\ast]$. According to Proposition 7.7 of \citet{Federico2023-pm}, this implies that $v''$ is locally bounded on $E$. Therefore, we can deduce from \eqref{J} that
		\begin{align}
			\widehat\cL v(x) + \zeta(x) &= 0 \;\;\textup{and} \;\;-\ell\leq v(x)\leq u &&\textup{if}\;\; \xL< x < \xU, \label{con}\\
			\widehat\cL v(x) + \zeta(x) &\leq 0 \;\;\textup{and} \;\;\phantom{-\ell<}v(x)=-\ell  &&\textup{if}\;\; x\leq\xL,\label{stop-l}\\
			\widehat\cL v(x) + \zeta(x) &\geq 0 \;\;\textup{and} \;\;\phantom{-\ell<}v(x)=u  &&\textup{if}\;\; x\geq\xU,\label{stop-u}
		\end{align}
		where 
		\begin{align*}
			\widehat\cL v(x) \triangleq \frac{1}{2}\sigma^2(x)v''(x) + (\alpha(x) + (\sigma'(x)-\theta^V(x))\sigma(x))v'(x) -\hat\rho(x)v(x),\;\;\;\text{on}\;C^2(E).
		\end{align*}
		The first equation in \eqref{con} comes from taking the derivative of $V$ with respect to $x$ over the intervals $(\xL,\xS)$. The first inequality in \eqref{stop-l} and \eqref{stop-u} follow from Condition 4 and 5 in Theorem~\ref{th:verification}, respectively.\\ 
		\indent Notice that $\widehat\cL$ is the characteristic operator on $C^2(E)$ of the killed process $\pro{e^{-\int_0^t \hat\rho(\hat X^{\theta^V}_s)\rd s}\hat X}$, which is well-defined since it is implied by Condition \eqref{con: loc lips}  that $\hat X$ also admits a strong unique solution. Therefore, by the variational inequalities of optimal stopping times (see, for example, \citet{Oksendal2010-da}, Theorem 10.4.1, or \citet{Oksendal2019}, Theorem 6.1), we can infer from \eqref{con} and \eqref{stop-l} that on $E$, $v(x)\leq  w(x,\tau^\ast,\eta)$ for any stopping time $\eta$. This means that $v(x)\leq \sup_{\tau} \inf_{\eta} w(x,\tau,\eta)$. On the other hand, \eqref{con} and \eqref{stop-l} induces $v(x)\geq  w(x,\tau,\eta^\ast)$, for any stopping time $\tau$, implying that $v(x)\geq  \inf_{\eta} \sup_{\tau} w(x,\tau,\eta)$. Since $\inf_{\eta} \sup_{\tau} w(x,\tau,\eta) \geq  \sup_{\tau} \inf_{\eta} w(x,\tau,\eta)$, we finally obtain \eqref{eq:dynkin}, which concludes the proof.
	\end{proof} 
	
	The following result is the refinement of Theorem~\ref{prop:dynkin}, which later used to examine the comparative static of ambiguity as well as giving an insight toward managerial concept of the Dynkin game in singular control.
	\begin{cor}\label{cor:dynkin-refine}
		Suppose that $v$ solves \eqref{eq:dynkin} and $(\ts,\es)$ is the saddle point \eqref{eq:saddle-point}. Then
		\begin{align}\label{eq:dynkin-refine}
			v(x)=\begin{cases}
				\underline{w}(x,\ts,\ps)&\text{if}\;\; x<\xS\\
				\overline{w}(x,\es,\ps)&\text{if}\;\; x\geq\xS
			\end{cases}
		\end{align}
		where
		\begin{align*}
			\underline{w}(x,\ts,\ps)&\triangleq\Ep^{\theta^V=+\kappa}_x\left[\int_0^{\ts\wedge \ps} e^{-\int_0^t\hat\rho(\hat X^{\theta^V}_s)\rd s}\zeta(\hat X^{\theta^V}_t)\rd t - 1_{(0,\ps)}(\ts)e^{-\int_0^{\ts}\hat\rho(\hat X^{\theta^V}_s)\rd s}\ell\right], \\
			\overline{w}(x,\es,\ps)&\triangleq\Ep^{\theta^V=-\kappa}_x\left[\int_0^{\es\wedge \ps} e^{-\int_0^t\hat\rho(\hat X^{\theta^V}_s)\rd s}\zeta(\hat X^{\theta^V}_t)\rd t + 1_{(0,\ps)}(\es)e^{-\int_0^{\es}\hat\rho(\hat X^{\theta^V}_s)\rd s}u\right]
		\end{align*}
		and $\ps\triangleq\inf\{t>0:\hat X^{\theta^V}_t=x^\ast\}$.
	\end{cor}
	\begin{proof}[Proof of Corollary~\ref{cor:dynkin-refine}]
		We provide details only for $\overline w$  since the proof of the representation of $\underline{w}$ is analogous. Given that $\xS<\xU$, by \eqref{con}, \eqref{stop-u}, and \eqref{eq:VM_x*-}-\eqref{eq:SP_x*} we have that $(v,\xS,\overline x)$ satisfy
		\begin{align*}
			\widehat\cL v(x) + \zeta(x) &= 0 \;\;\;\;\textup{if}\;\; \xS< x < \xU, \\
			v(x) &= u \;\;\;\;\textup{if}\;\; x\geq\xU,\\
			v(\xS) &= 0 \;\;
		\end{align*}
		Since $v\in C^1([\xS,\infty])\cap C^2([\xS,\infty]\setminus \{\xU\})$, an application of Dynkin's formula yields that, for any $x\geq\xS,$
		\begin{align*}
			v(x)=\Ep^{\theta^V}_x\left[\int_0^{\es\wedge \ps} e^{-\int_0^t\hat\rho(\hat X^{\theta^V}_s)\rd s}\zeta(\hat X^{\theta^V}_t)\rd t + 1_{(0,\ps)}(\es)e^{-\int_0^{\es}\hat\rho(\hat X^{\theta^V}_s)\rd s}u\right].
		\end{align*}
		The claim representation of $\overline{w}$ follows by noticing that $\theta^V(x)=-\kappa$ for any $x\geq\xS$.
	\end{proof}
	\begin{remark}
		In managerial terms, Corollary~\ref{cor:dynkin-refine} can be interpreted as two timing games, one on $(-\infty, \xS)$ and the other on $[\xS, \infty)$, involving two players: the DM and a hypothetical player (nature). On the interval $(-\infty, \xS)$, the game requires the DM and nature selecting stopping times $\tau$ and $\phi$, respectively. The game ends when time reaches $\tau\wedge\phi$, at which point the DM receives a payoff of $\underline{w}(x,\tau,\phi)$ from nature. The DM's objective is to choose $\tau$ to maximize $\underline{w}(x,\tau,\phi)$, while nature select $\phi$ to minimize its payment.\\ \indent
		On the other hand, if the inventory level lies within $[\xS, \infty)$, the DM and nature must choose stopping times $\eta$ and $\phi$, respectively. When the game concludes at $\eta\wedge\phi$, the DM pays $\overline{w}(x,\es,\ps)$ to nature. Thus, the DM's goal is to minimize the payment by selecting $\eta$, while nature seeks to maximize its payoff by determining $\phi$.\\ \indent
		From~\eqref{eq:dynkin-refine}, it follows that the DM's ideal payoff occurs when $\underline{w}(\xS,\ts,\ps)=\overline{w}(\xS,\es,\ps) = 0$. In essence, the DM's optimal strategy is to keep the running marginal cost of holding inventory as close to zero as possible under timing games against the uncertain nature.
	\end{remark}
	
	\section{Comparative statics with Arithmetic Brownian Motion}\label{sec:ABM}
	Suppose that the uncontrolled cash inventory $X^0$ follows, under $\Pp_x$, the ABM
	\begin{equation}\label{X abm}
		X^0_t = x+\alpha t+\sigma B_t,
	\end{equation}
	so that
	\begin{equation}\label{fun sol abm}
		\vH_{\pm\kappa}(x) = e^{\beta_{\pm\kappa}x}, \quad\text{and}\quad 
		\vC_{\pm\kappa}(x) = e^{\gamma_{\pm\kappa}x},
	\end{equation}
	where $\beta_{\pm\kappa}>0$ and $\gamma_{\pm\kappa}<0$ are the positive and negative roots, respectively, of the quadratic equation
	\begin{equation}
		\cQ_{\pm\kappa}(\chi)\equiv \frac{1}{2}\sigma^2\chi^2+\left(\alpha\pm\kappa\sigma\right)\chi-\rho=0.
	\end{equation}
	Recall that the holding costs of cash are given by
	\begin{equation}
		c(x) = \ch  x\cdot 1_{[0,\infty)}(x)-\cc x\cdot 1_{(-\infty,0)}(x),
	\end{equation}
	for some $\ch  ,\cc >0$. As mentioned before, if $\xL<0<\xU$, then Conditions~4 and~5 in Theorem~\ref{th:verification} reduce to
	\begin{equation*}
		\cc \geq \rho\ell,\quad\text{and}\quad \ch  \geq \rho u,
	\end{equation*}
	i.e. the control costs must not exceed the expected discounted uncontrolled holding costs; otherwise, it would never be optimal to exercise control.
	
	The expected discounted uncontrolled holding costs in this case are given by
	\begin{equation*}
		R_{\pm\kappa}(x) =
		\begin{cases}
			-\frac{\cc }{\rho}\left[x+\frac{\alpha\pm\kappa\sigma}{\rho}\right] +\hat{E}_{\pm\kappa}e^{\beta_{\pm\kappa}x} &\text{if $x<0$}\\
			+\frac{\ch  }{\rho}\left[x+\frac{\alpha\pm\kappa\sigma}{\rho}\right] +\check{E}_{\pm\kappa}e^{\gamma_{\pm\kappa}x} &\text{if $x\geq 0$}
		\end{cases},
	\end{equation*}
	where
	\begin{align*}
		&\hat{E}_{\pm\kappa} = \frac{\ch  +\cc }{\rho^2}\frac{\rho-\gamma_{\pm\kappa}(\alpha\pm\sigma\kappa)}{\beta_{\pm\kappa}-\gamma_{\pm\kappa}},
		\quad\text{and}\quad
		\\&\check{E}_{\pm\kappa} = \frac{\ch  +\cc }{\rho^2}\frac{\rho-\beta_{\pm\kappa}(\alpha\pm\sigma\kappa)}{\beta_{\pm\kappa}-\gamma_{\pm\kappa}}.
	\end{align*}
	Since 
	\begin{equation*}
		\cQ_{\pm\kappa}(\beta_{\pm\kappa})=\cQ_{\pm\kappa}(\gamma_{\pm\kappa})=0,
	\end{equation*}
	the constants $\hat{E}$ and $\check{E}$ can be written as
	\begin{align*}
		&\hat{E} = \frac{\ch  +\cc }{2\rho^2}\frac{\sigma^2\gamma_{\pm\kappa}^2}{\beta_{\pm\kappa}-\gamma_{\pm\kappa}}>0,
		\quad\text{and}\quad
		\\&\check{E} = \frac{\ch  +\cc }{2\rho^2}\frac{\sigma^2\beta_{\pm\kappa}^2}{\beta_{\pm\kappa}-\gamma_{\pm\kappa}}>0,
	\end{align*}
	respectively.
	That is, the expected discounted holding costs of uncontrolled cash inventory equals
	\begin{equation}\label{R abm}
		R_{\pm\kappa}(x) =
		\begin{cases}
			\frac{-\cc }{\rho}\left[x+\frac{\alpha\pm\kappa\sigma}{\rho}\right] +\frac{(\ch  +\cc )\sigma^2\gamma^2_{\pm\kappa}}{2\rho^2(\beta_{\pm\kappa}-\gamma_{\pm\kappa})}e^{\beta_{\pm\kappa}x} &\text{if $x<0$}\\
			\frac{+\ch  }{\rho}\left[x+\frac{\alpha\pm\kappa\sigma}{\rho}\right] +\frac{(\ch  +\cc )\sigma^2\beta^2_{\pm\kappa}}{2\rho^2(\beta_{\pm\kappa}-\gamma_{\pm\kappa})}e^{\gamma_{\pm\kappa}x} &\text{if $x\geq 0$}
		\end{cases}.
	\end{equation}
	We recall that the general solution of the value function takes the form of \eqref{J}. Therefore, in the case an arithmetic Brownian motion, substituting \eqref{fun sol abm} and \eqref{R abm} into \eqref{J} reads
	\begin{equation}\label{J abm}
		V(x) =  
		\begin{cases}
			\ell|\xL-x| -\overbrace{\frac{1}{\rho}\big[\ell(\alpha -\kappa\sigma) +\cc \xL\big]}^{\triangleq \Gamma^-(\xL)} &\text{if    $x\leq\xL$}\\
			R_{-\kappa}(x)+A\vH_{-\kappa}(x)+B\vC_{-\kappa}(x) &\text{if   } \xL<x<x^\ast\\
			R_{+\kappa}(x)+C\vH_{+\kappa}(x)+D\vC_{+\kappa}(x) &\text{if   } x^\ast\leq x<\xU\\
			u|x-\xU|+\underbrace{\frac{1}{\rho}\big[u(\alpha +\kappa\sigma) +\ch  \xU\big]}_{\triangleq \Gamma^+(\xU)} &\text{if   } x\geq\xU.
		\end{cases}
	\end{equation}
	for some constant $A,B,C$ and $D$. 
	
	In the following, we perform a sensitivity analysis of the optimal control boundaries $\xL$ and $\xU$ with respect to same model parameters. 
	
	To keep the argument concise, we assume from this point onward that Assumption~\eqref{ass:X}
	holds, that $X^0$ satisfies \eqref{X abm}, and that $(\xL, \xU)$ is a solution to \eqref{J abm}.
	\begin{proposition}[Comparative statics of risk]\label{prop:com risk abm}
		The control barriers $(\xL,\xU)$ expand as $\sigma$ increases. That is, $\sigma\mapsto\xL(\sigma)$ is non-increasing and $\sigma\mapsto\xU(\sigma)$ is non-decreasing.
	\end{proposition}
	\begin{proof}[Proof of Proposition~\ref{prop:com risk abm}]
		The idea of the proof is adopted from \citet{Matomaki2012-vy}, Theorem 6.1 (also shown in \citet{Ferrari2020-oh}.) Suppose that $\sigma_1$, where $\sigma\leq\sigma_1$, is a diffusion term of the uncontrolled arithmetic Brownian motion $X^{0,\sigma_1}_t= x +\alpha t +\sigma_1 B_t$. Let $\cL_{\sigma_1}$ be the characteristic operator on $C^2(E)$ of the killed process $\pro{e^{-\rho t}X^{0,\sigma_1}}$, that is, $\cL_{\sigma_1}\varphi(x)\triangleq\frac{1}{2}\sigma_1^2\varphi''(x)+\alpha\varphi'(x)-\rho\varphi(x)$, and  $V_{\sigma_1}(x)$ is the associated value function. 
		Then a straightforward calculation gives
		\begin{align}\label{L sigma 1}
			\cL_{\sigma_1}V(x)-\theta^\ast\sigma_1 V'(x) +c(x)= \begin{cases}
				\Gamma^-(\xL) - \Gamma^-(x) &\textup{if}\;\;\; x\leq\xL\\
				\frac{1}{2}(\sigma_1^2 - \sigma^2)V''(x) -\kappa(\sigma_1-\sigma)V'(x)  &\textup{if}\;\;\; \xL<x<\xS\\
				\frac{1}{2}(\sigma_1^2 - \sigma^2)V''(x) +\kappa(\sigma_1-\sigma)V'(x)  &\textup{if}\;\;\; \xS\leq x<\xU\\
				\Gamma^+(x) - \Gamma^+(\xU) &\textup{if}\;\;\; x\geq\xU.
			\end{cases}
		\end{align}
		One can see that $\cL_{\sigma_1}V(x)-\theta^\ast\sigma_1 V'(x) +c(x)\geq 0$ for any $x\in E$. This is because $\Gamma^-(\xL) - \Gamma^-(x)\geq 0$ if $x\leq\xL$ and also $\Gamma^+(x) - \Gamma^+(\xU)\geq 0$ if $x\geq\xU$. The claim holds on $(\xL,\xU)$ because $V$ is convex is $x$, thanks to Theorem~\ref{th:verification2}, and the fact that $\xS$ situates the point where worst-case changes, i.e. $V'(x)\leq 0$ if $x<\xS$ and $V'(x)\geq 0$, otherwise. Since $V_{\sigma_1}(x)$ is the value function when $\sigma = \sigma_1$, Theorem~\ref{th:verification} asserts that 
		\begin{align}
			\cL_{\sigma_1}V(x)-\theta^\ast\sigma_1 V'(x) +c(x) \geq \cL_{\sigma_1}V_{\sigma_1}(x)-\theta^\ast\sigma_1 V_{\sigma_1}'(x) +c(x)=0
		\end{align}
		Since this inequality holds for any $\sigma_1\geq\sigma$, it is implied by It\=o's lemma that $V_{\sigma_1}(x)\geq V(x)$ for all $x\in E$. This completes the first part of the proof.
		\\ \indent Now, we demonstrate that $(\xL, \xU)$ expands as $\sigma$ increases. To do this, we assume that $\sigma_1 > \sigma\triangleq\sigma_0$ and denote $(\xL_1, \xU_1)$ as the optimal control barriers of $V_{\sigma_1}(x)$. If we assume in contrary that $\xL_1 > \xL \triangleq \xL_0$ and $\xU_1 < \xU \triangleq \xU_0$, then Condition~4 in Theorem~\ref{th:verification} gives
		\begin{align*}
			\cc (\xL_i-x)\geq\ell\big[\rho(\xL_i-x)-\big(\big(\alpha(\xL_i)-\alpha(x)\big)-\kappa\big(\sigma_i(\xL_i)-\sigma_i(x)\big)\big],\;\;\textup{for}\;\;x\leq \xL_0<\xL_1,\;i=0,1,
		\end{align*}
		implying that
		\begin{align*}
			V_{\sigma_0}(\xL_0) = -\frac{1}{\rho}(\ell(\alpha-\kappa\sigma_0)+\cc  \xL_0) > \ell(\xL_1-\xL_0) -\frac{1}{\rho}(\ell(\alpha-\kappa\sigma_1)+\cc  \xL_1)=V_{\sigma_1}(\xL_0),
		\end{align*}
		which is a contradiction. Therefore, we conclude that $\xL_1 \leq \xL_0 $. A similar argument also gives that  $\xU_1 \geq \xU_0 $. Hence, these result in an expansion of the control barriers as $\sigma$ increases, concluding the proof.
	\end{proof}
	
\begin{figure}[!htb]
		\begin{subfigure}{0.5\textwidth}
			\centering
			\includegraphics[width=\linewidth]{ 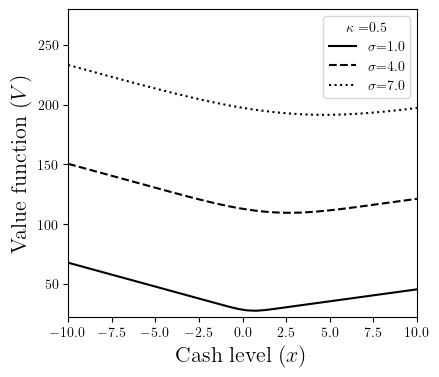}
			\caption{Value functions given $\kappa=0.5$ for $\sigma\in\{1,4,7\}$.}\label{fig:VF_sigma}
		\end{subfigure}\hfill
		\begin{subfigure}{0.5\textwidth}
			\centering
			\includegraphics[width=\linewidth]{ 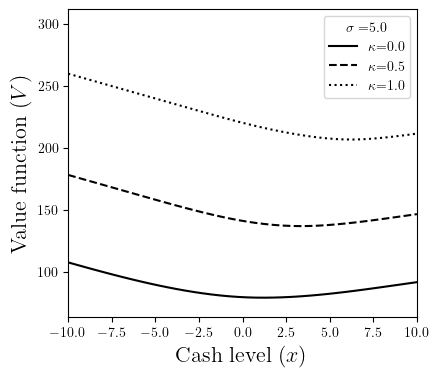}
			\caption{Value functions given $\sigma=5$ for $\kappa\in\{0,0.5,1\}$.}\label{fig:VF_kappa}
		\end{subfigure}
		\caption{Value functions with different levels of (\subref{fig:VF_sigma}) risk and (\subref{fig:VF_kappa}) ambiguity, given sample parameters $\alpha=0$, $\rho=0.1$, $u=3$, $\ell=5$, and $\check{c}=\hat{c}=1$.}
	\end{figure}

	Not surprisingly, this aligns with the familiar result that greater risk increases the firm’s cost function. The same applies to ambiguity, as becomes evident by extending the first part of the proof of Proposition~\ref{prop:com risk abm} to account for a higher level of $\kappa$. These can be depicted by Figure~\ref{fig:VF_sigma} and~\ref{fig:VF_kappa}. Specifically, a manager with maxmin utility assigns a higher expected cost to inventory management under such conditions. Furthermore, the expansion of the control barriers suggests that increased risk causes delays in taking action. This occurs because a rise in $\sigma$ makes extreme scenarios more probable, meaning it becomes, on average, just as likely for the inventory to reach the target level as it is to trigger a control action. Consequently, the optimal singular control policy indicates that acting too frequently becomes increasingly costly.
	It is evident that Conditions 4 and 5 in Theorem~\ref{th:verification} are essential for the previous result. However, these conditions alone are not enough to offer a comparative static analysis of ambiguity (or of the other model parameters). This is where the Dynkin game of Theorem~\ref{prop:dynkin} becomes relevant and is therefore utilized to support the following arguments. 
	
	It is important to note that, when ambiguity arises, the comparative statics may exhibit non-monotonicity, depending on the \emph{symmetry} of the model parameters. This phenomenon does not appears in the standard comparative statics of two-sided singular control, as discussed in \citet{Matomaki2012-vy} and \citet{Ferrari2020-oh}, for instance. To explore this, we begin by introducing the notion of symmetry as defined below.
	\begin{definition}
		Suppose that $\alpha,\rho,\kappa,\sigma,\hat c,\check c,\ell$ and $u$ satisfy Condition 4 and 5 in Theorem \ref{th:verification}. Then the model parameters are said to be
		\begin{itemize}
			\item (symmetric) if $\alpha=0,\check c = \hat c$ and $\ell=u$,
			\item (asymmetric) if $\alpha\neq0$, $\check c \neq \hat c$ or $\ell\neq u$.
		\end{itemize}
	\end{definition}
	The following lemma will be useful for our comparative statics arguments.
	\begin{lemma}\label{lem:symmetry}
		Suppose that the model parameters are symmetric, i.e. $\alpha=0$, $u=\ell$, and $\hat{c}=\check{c}$. Then $\xS=0$, $|\xL|=|\xU|$, and the marginal value function is an odd function: $v(x)=-v(-x)$ for all $x\in E$.
	\end{lemma}
	\begin{proof}[Prove of Lemma~\ref{lem:symmetry}]
		Suppose, ad absurdum, that $\xS>0$ and $|\xL|=|\xU|$. Adapting the argument in \citet[Section~3.3]{Michael_Harrison2013-fe}, it holds that,
		\begin{align}
			\Pp^{\theta^V}(\ps<\ts|\hat X^{\theta^V}_0=x<\xS)&=\begin{cases}
				\frac{x-\xL}{\xS-\xL}&\;\;\text{if}\;\;\kappa=0\\
				\frac{ e^{\frac{2\kappa}{\sigma}x} - e^{\frac{2\kappa}{\sigma}\xL}}{e^{\frac{2\kappa}{\sigma}\xS} - e^{\frac{2\kappa}{\sigma}\xL}},&\;\;\text{if}\;\;\kappa>0
			\end{cases}\label{eq:P phi tau}
			\\
			\Pp^{\theta^V}(\ps<\es|\hat X^{\theta^V}_0=x>\xS)&=\begin{cases}
				\frac{x-\xU}{\xS-\xU}&\;\;\text{if}\;\;\kappa=0\\
				\frac{ e^{-\frac{2\kappa}{\sigma}x} - e^{-\frac{2\kappa}{\sigma}\xU}}{e^{-\frac{2\kappa}{\sigma}\xS} - e^{-\frac{2\kappa}{\sigma}\xU}},&\;\;\text{if}\;\;\kappa>0
			\end{cases}\label{eq:P phi eta}
		\end{align}
		respectively. Now suppose that $\underline\phi \triangleq\inf\{t>0:\hat X^{\theta^V}_t=-\xS\} $. If $\xS>0$, then for any $\delta>0$ such that $\xL+\delta\leq \xS$ and $\xU-\delta\geq\xS$, we have
		\begin{align*}
			\frac{\xU-\delta - \xU}{\xS-\xU}=\frac{-\delta }{\xS+\xL}&>\frac{-\delta }{-\xS+\xL}=\frac{\xL+\delta -\xL }{\xS-\xL}&&\text{if}\;\;\kappa=0,\;\;\text{and}\\
			\frac{ e^{-\frac{2\kappa}{\sigma}(\xU-\delta)} - e^{-\frac{2\kappa}{\sigma}\xU}}{e^{-\frac{2\kappa}{\sigma}\xS} - e^{-\frac{2\kappa}{\sigma}\xU}} &> \frac{ e^{\frac{2\kappa}{\sigma}(\xL +\delta)} - e^{\frac{2\kappa}{\sigma}\xL}}{e^{\frac{2\kappa}{\sigma}\xS} - e^{\frac{2\kappa}{\sigma}\xL}} &&\text{if}\;\;\kappa>0,
		\end{align*}
		which implies that 
		\begin{equation}\label{eq:hitting}
			\Pp^{\theta^V}_{\xU-\delta}(\ps<\es) > \Pp^{\theta^V}_{\xL +\delta}(\ps<\ts).
		\end{equation} \indent
		Suppose now that $\hat X^{-\kappa}_0 = x$ and $\hat X^{+\kappa}_0=-x$, with $x>\xS\,(>0)$. Then it is more likely that $\hat X^{\theta^V}_0$ starting from above zero arrives at $\xS$ earlier. Consequently, it follows from Corollary \ref{cor:dynkin-refine} that
		\begin{align*}
			\underline{w}(\xL+\delta)&= \Ep^{\theta^V}_{\xL+\delta}\left[\int_0^{\ts\wedge \ps} e^{-\rho t}(\ch  \cdot 1_{(0,\infty)}(\hat X^{\theta^V}_t)-\cc \cdot 1_{(-\infty,0)}(\hat X^{\theta^V}_t))\rd t - 1_{(0,\ps)}(\ts)e^{-\rho \ts}\ell\right]\\
			&=-\Ep^{\theta^V}_{\xL+\delta}\left[\int_0^{\ts\wedge \ps} e^{-\rho t}(\ch  \cdot 1_{(-\infty,0)}(\hat X^{\theta^V}_t)-\cc \cdot 1_{(0,\infty)}(\hat X^{\theta^V}_t))\rd t + 1_{(0,\ps)}(\ts)e^{-\rho \ts}u\right]\\
			&\underset{\eqref{eq:hitting}}{<} -\Ep^{\theta^V}_{\xU-\delta}\left[\int_0^{\es\wedge \ps} e^{-\rho t}(\ch  \cdot 1_{(0,\infty)}(\hat X^{\theta^V}_t)-\cc \cdot 1_{(-\infty,0)}(\hat X^{\theta^V}_t))\rd t + 1_{(0,\ps)}(\es)e^{-\rho \es}u\right]\\
			&=-\overline{w}(\xU-\delta).
		\end{align*}
		Note that the inequality holds by \eqref{eq:hitting} alone. This is due to symmetry in the marginal holding cost around zero, i.e. 
		\begin{align*}
			\Pp^{\theta^V}_{\xL-\delta}(\hat X^{\theta^V}_t<0)=\Phi\left(\frac{-(\xL-\delta)-(-\kappa\sigma)t}{\sigma\sqrt{t}}\right)=\Phi\left(\frac{(\xU+\delta)+\kappa\sigma t}{\sigma\sqrt{t}}\right)=\Pp^{\theta^V}_{\xU+\delta}(\hat X^{\theta^V}_t>0),
		\end{align*}
		where $\Phi$ is cumulative distribution function of the standard normal distribution, and that $\cc=\ch$. Therefore, we can imply that $\ell=-\underline{w}(\xL)>\overline{w}(\xU)=u$, which is a contradiction. A similar technique can be easily adopted to examine a contradiction when $\xS=0$ and $|\xL|\neq|\xU|$, so omitted for brevity. This completes the proof. 
	\end{proof}
	
	Using Lemma~\ref{lem:symmetry} we can derive several comparative statics results, starting from a symmetric base case. First we show that the continuation region $(\xL,\xU)$ ``shrinks'' as $\kappa$ increases.
	\begin{proposition}[Comparative static of ambiguity]\label{prop:com amb abm}
		Suppose that the model parameters are symmetric. Then, the control barriers $(\xL,\xU)$ shrink as $\kappa$ increases. That is, $\kappa\mapsto\xL(\kappa)$ is non-decreasing and $\kappa\mapsto\xU(\kappa)$ is non-increasing.
	\end{proposition}
	\begin{proof}[Prove of Proposition~\ref{prop:com amb abm}]
		Suppose that $\kappa_1$, where $\kappa_0\triangleq\kappa\leq\kappa_1$, and that denoted by $v_{\kappa_i}(x)$ is the first derivative with respect to $x$ of the value function $V(x;\kappa_i)$, and $\theta^V_{i}\triangleq \kappa_i1_{(-\infty,\xS)}(x)-\kappa_i1_{[\xS,\infty)}(x)$ for each $\kappa_i,i=0,1$.  According to the comparison theorem for It\=o's processes (\citet{karatzas1991brownian}, Chapter 5.2, or \citet{Protter2010-yj}, Theorem 52), together with \eqref{def:thetaV}, one can see that 
		\begin{align*}
			\hat X^{\theta^V_0}_t&\geq\hat  X^{\theta^V_1}_t  ,\;\;\text{on }(-\infty,\xS),\;\;\text{and}\;\;
			\hat X^{\theta^V_1}_t\geq\hat  X^{\theta^V_0}_t,\;\;\text{on }[\xS,\infty).
		\end{align*}
		$\Pp_x$-a.s. Since $\zeta(x)$ is non-decreasing in $x$ and $\alpha'(x)=\sigma'(x)=0$ for all $x\in E$, we have by Theorem~\ref{prop:dynkin} and the comparison theorem that on $(-\infty,\xS)$,
		\begin{align}\label{eq:v kappa}
			v_{\kappa_0}(x)=V'(x)&=w_{\kappa_0}(x,\ts,\ps)\nonumber\\
			&=
			\Ep^{\theta^V_0}_x\Big[\int_0^{\ts\wedge\ps }e^{-\rho t}\zeta(\hat X^{\theta^V_0}_t)\rd t- 1_{(0,\ps ]}(\ts )e^{-\rho\ts}\ell  \Big]\nonumber\\
			&=\Ep_x\Big[\int_0^{\ts\wedge\ps }e^{-\rho t}\zeta(\hat X^{0}_t-\kappa_0\sigma t)\rd t- 1_{(0,\ps ]}(\ts )e^{-\rho\ts}\ell  \Big]\nonumber\\
			&\geq\Ep_x\Big[\int_0^{\ts\wedge\ps }e^{-\rho t}\zeta(\hat X^{0}_t-\kappa_1\sigma t)\rd t- 1_{(0,\ps ]}(\ts )e^{-\rho\ts}\ell  \Big]\nonumber\\
			&= \Ep^{\theta^V_1}_x\Big[\int_0^{\ts\wedge\ps }e^{-\rho t}\zeta(\hat X^{\theta^V_1}_t)\rd t- 1_{(0,\ps ]}(\ts )e^{-\rho\ts}\ell  \Big]\nonumber\\
			&=w_{\kappa_1}(x,\ts,\ps)=v_{\kappa_1}(x).
		\end{align}
		Since this result holds for any $x<\xS$, it follows that $-\ell=v_{\kappa_0}(\xL)\geq v_{\kappa_1}(\xL).$ Because $V_{\kappa_1}(x)$ is convex on $E$, we know that there exists $\delta>0$ such that $ v_{\kappa_1}(\xL+\delta) = -\ell$. This means that $\xL$ increases as $\kappa$ increases. Since the parameters are symmetric, it follows from Lemma~\ref{lem:symmetry} that $v_{\kappa_1}(\xL+\delta)=-v_{\kappa_1}(\xU-\delta)=-u$, implying that $\xU$ decreases as $\kappa$ increases. Following a similar procedure, it is easy to show that the same result holds when $x>\xS$. In total, we conclude that $(\xL,\xU)$ shrinks as the level of ambiguity rises.
	\end{proof}
	
	\begin{figure}
		\centering
		\includegraphics[width=0.45\textwidth]{ 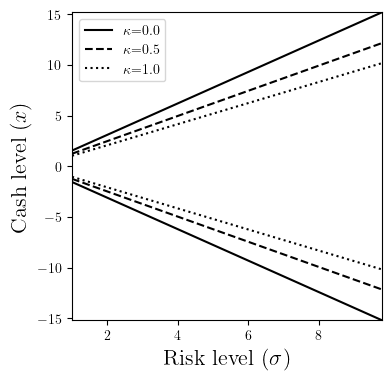}
		\caption{Control barriers as a function of $\sigma$ in the symmetric case, using base-case parameters: $\alpha = 0$, $\rho = 0.1$, $\ell = u = 2$, and $\cc = \ch = 1$. Each line style—solid, dashed, and dotted—represents control barriers for $\kappa \in \{0, 0.5, 1\}$, respectively. The upper line corresponds to the upper control barrier, $\xU$, while the lower line represents the lower control barrier, $\xL$. Given a fixed level of risk, when the cash level reaches $\xU$ ($\xL$), the DM must intervene to maintain the cash level below $\xU$ (above $\xL$), incurring a proportional cost $u$ ($\ell$). } 
		\label{fig:CB_sym}
	\end{figure}
	
	Proposition~\ref{prop:com amb abm} implies that a more ambiguous DM exerts control earlier and keeps a smaller cash inventory. This contrasts with the comparative  statics of risk and is illustrated by Figure~\ref{fig:CB_sym}. This results reflects the pessimistic mindset of managers who lack full confidence in the drift of the inventory flow process. Mathematically speaking, $\kappa$-ignorance increases the drift when the value function is increasing and decreases it when the value function is decreasing. As a result, the manager faces higher expected holding costs near the control barriers. Thus, the optimal control policy implies that taking action earlier becomes less costly. However, this inevitably leads to a higher average frequency of control exertion, which, as previously discussed, contributes to an increase in the value function.
	
	Next, we conduct a comparative static analysis for the remaining parameters under a fixed level of ambiguity $\kappa$. For simplicity, we start from the symmetric base case. We start by looking at changes in the discount rate. 
	
	\begin{proposition}[Comparative statics of discounted rate]\label{prop:com disc abm} Suppose that the model parameters are symmetric. Then, the control barriers $(\xL,\xU)$ expand as $\rho$ increases. That is, $\rho\mapsto\xL(\rho)$ is non-increasing and $\rho\mapsto\xU(\rho)$ is non-decreasing.
	\end{proposition}
	\begin{proof}[Proof of Proposition~\ref{prop:com disc abm}]
		Suppose that $\rho_1$ where $\rho_0\triangleq\rho\leq\rho_1$ and denoted by $v_{\rho_i}(x)$ is the first derivative with respect to $x$ of the value function $ V(x;\rho_i)$, for each $\rho_i,i=0,1$. Applying the It\={o}'s integration by parts to~\eqref{eq:dynkin} yields
		\begin{align*}
			v_{\rho_0}(x) \triangleq v(x)&=\Ep^{\theta^V}_x\Big[\int_0^{\ts\wedge\ps }e^{-\rho_0 t}\zeta(\hat X^{\theta^V}_t)\rd t- 1_{(0,\ps ]}(\ts )e^{-\rho_0\ts}\ell  \Big]\\
			&=\Ep^{\theta^V}_x\Big[\int_0^{\ts\wedge\ps }(-\rho_0 v(\hat X^{\theta^V}_t)+\zeta(\hat X^{\theta^V}_t))\rd t- 1_{(0,\ps ]}(\ts )\ell  \Big]\\
			&\leq
			\Ep^{\theta^V}_x\Big[\int_0^{\ts\wedge\ps }(-\rho_1 v(\hat X^{\theta^V}_t)+\zeta(\hat X^{\theta^V}_t))\rd t- 1_{(0,\ps ]}(\ts )\ell  \Big]\\
			&=
			\Ep^{\theta^V}_x\Big[\int_0^{\ts\wedge\ps }e^{-\rho_1 t}\zeta(\hat X^{\theta^V}_t)\rd t- 1_{(0,\ps ]}(\ts )e^{-\rho_1\ts}\ell  \Big]\\
			&=v_{\rho_1}(x)
		\end{align*}
		for any $x\leq\xS$. The inequality holds because $v(x)\leq0$ for all $x\leq\xS$. In turn, this implies that $-\ell=v_{\rho_0}(\xL)\leq v_{\rho_1}(\xL)$. Since $v_{\rho_1}(x)$ is increasing in $x$, we can deduce that there exists $\delta>0$ such that $v_{\rho_1}(\xL-\delta)=-\ell$, meaning that $\xL$ decreases as $\rho$ increases. By the symmetry assumption, we deduce that $v_{\rho_1}(\xL-\delta)=v_{\rho_1}(\xU+\delta)=-u$, that is, $\xU$ increases as $\rho$ increases. This concludes that $(\xL,\xU)$ expands as $\rho$ increases.
	\end{proof}
	
	This result implies that a more patient DM exerts less control. This happens because, from the DM's perspective, the worst-case prior is less bad, because expected future payments are discounted more.
	
	The effect of a change in the reference drift is asymmetric: the inaction region shifts downwards.  
	
	\begin{proposition}[Comparative statics of drift]\label{prop:com drift abm} Suppose that the model parameters are symmetric. Then the control barriers $(\xL,\xU)$ shift downward as $\alpha$ increases. That is, $\alpha\mapsto\xL(\rho)$ and $\alpha\mapsto\xU(\rho)$ are non-increasing.
	\end{proposition}
	\begin{proof}[Proof of Proposition~\ref{prop:com drift abm}]
		Suppose that $\alpha_1$ where $\alpha_0\triangleq\alpha\leq\alpha_1$ and denoted by $v_{\alpha_i}(x)$ is the first derivative with respect to $x$ of the value function $ V(x;\alpha_i)$, for each $\alpha_i,i=0,1$. 
		Given that $\zeta(x)$ is non-decreasing in $x$ and $\alpha'(x) = \sigma'(x) = 0$ for all $x \in E$, Theorem~\ref{prop:dynkin} implies that
		\begin{align*}
			v_{\alpha_0}(x)&=\Ep^{\theta^V}_x\Big[\int_0^{\ts\wedge\es }e^{-\rho t}\zeta(\hat X^{\theta^V}_t)\rd t- 1_{(0,\es)}(\ts)e^{-\rho\ts}\ell + 1_{(0,\ts)}(\es)e^{-\rho\es}u\Big]\\
			&\leq\Ep^{\theta^V}_x\Big[\int_0^{\ts\wedge\es }e^{-\rho t}\zeta(\hat X^{\theta^V}_t+ (\alpha_1-\alpha_0)t)\rd t- 1_{(0,\es)}(\ts)e^{-\rho\ts}\ell+ 1_{(0,\ts)}(\es)e^{-\rho\es}u\Big]\\
			&=v_{\alpha_1}(x)
		\end{align*}
		for any $x\in E$. In turn, this implies that $-\ell=v_{\alpha_0}(\xL)\leq v_{\alpha_1}(\xL)$ and $u=v_{\alpha_0}(\xU)\leq v_{\alpha_1}(\xU)$. Since $v_{\alpha_1}(x)$ is increasing in $x$, we can deduce that there exists $\delta,\delta'>0$ such that $v_{\alpha_1}(\xL-\delta)=-\ell$ and $v_{\alpha_1}(\xU-\delta')=u$, meaning that $(\xL,\xU)$ shift downward as $\alpha$ increases.
	\end{proof}
	
	The intuition behind this result is as follows. An increasing reference drift makes it more likely, under the worst-case measure, that holding costs $\hat{c}$ need to be paid. That makes it more attractive to control the inventory on the upside, leading to a decrease in $\xU$. Conversely, with higher $\alpha$ it is less likely that holding costs $\check{c}$ are incurred, which leads the DM to reduce the control barrier $\xL$. The comparative statics for the holding costs follow a similar pattern.
	
	\begin{proposition}[Comparative statics of holding costs]\label{prop:com hold cost abm}Suppose that the model parameters are symmetric. Then the control barriers $(\xL,\xU)$ shift upward as $\cc $ increases, but shift downward as $\ch$ increases. That is, $\cc \mapsto\xL(\cc )$ and $\cc \mapsto\xU(\cc )$ are non-decreasing, while $\ch  \mapsto\xL(\ch  )$ and $\ch  \mapsto\xU(\ch  )$ are non-increasing.
	\end{proposition}
	\begin{proof}[Proof of Proposition~\ref{prop:com hold cost abm}]
		Suppose that $\cc_1$ where $\cc_0\triangleq\cc\leq\cc_1$ and denoted by $v_{\cc_i}(x)$ is the first derivative with respect to $x$ of the value function $ V(x;\cc_i)$, for each $\cc_i,i=0,1$. Then it follows from Theorem~\ref{prop:dynkin} that 
		\begin{align*}
			v_{\cc_0}(x) &= \Ep^{\theta^V}_x\Big[\int_0^{\ts\wedge\es }e^{-\rho t}(\ch  \cdot 1_{(0,\infty)}(\hat X^{\theta^V}_t)-\cc_0 \cdot 1_{(-\infty,0)}(\hat X^{\theta^V}_t))\rd t \\ &\hspace{5cm}- 1_{(0,\es ]}(\ts )e^{-\rho\ts}\ell  + 1_{(0,\ts)}(\es)e^{-\rho\es}u\Big]\\
			&\geq \Ep^{\theta^V}_x\Big[\int_0^{\ts\wedge\es }e^{-\rho t}(\ch  \cdot 1_{(0,\infty)}(\hat X^{\theta^V}_t)-\cc_1 \cdot 1_{(-\infty,0)}(\hat X^{\theta^V}_t))\rd t \\ &\hspace{5cm}- 1_{(0,\es ]}(\ts )e^{-\rho\ts}\ell  + 1_{(0,\ts)}(\es)e^{-\rho\es}u\Big]\\
			&= v_{\cc_1}(x),
		\end{align*}
		for all $x\in E$, which implies that  $-\ell=v_{\cc_0}(\xL)\geq v_{\cc_1} (\xL)$ and $u=v_{\cc_0}(\xU)\geq v_{\cc_1} (\xU)$. As a result, there is $\delta,\delta'>0$ such that $v_{\cc_1}(\xL+\delta)=-\ell$ and $v_{\cc_1}(\xU+\delta')=u$, since $v_{\cc_1}(x)$ does not decrease in $x$. Therefore, $(\xL,\xU)$ shift upward as $\cc$ increases. One can use the same argument to show that $(\xL,\xU)$ shift downward as $\ch$ increases. 
	\end{proof}
	
	\begin{proposition}[Comparative statics of control costs]\label{prop:com control cost abm} Suppose that the model parameters are symmetric. Then the control barriers $(\xL,\xU)$ expand as $\ell$ or $u$ increase. That is, $(\ell,u) \mapsto\xL(\ell,u)$ is non-increasing, while $(\ell,u) \mapsto\xU(\ell,u)$ is non-decreasing.
	\end{proposition}
	\begin{proof}[Proof of Proposition~\ref{prop:com control cost abm}]
		Suppose that $\ell_1$ where $\ell_0\triangleq\ell\leq\ell_1$ and denoted by $v_{\ell_i}(x)$ is the first derivative with respect to $x$ of the value function $ V(x;\ell_i)$, for each $\ell_i,i=0,1$. If $x\leq\xS$, then it follows from Corollary~\ref{cor:dynkin-refine} that 
		\begin{align*}
			v_{\ell_0}(x) +\ell_0 &=\Ep^{\theta^V}_x\Big[\int_0^{\ts\wedge\ps }e^{-\rho t}\zeta(\hat X^{\theta^V}_t)\rd t+ 1_{(0,\ps ]}(\ts )(1-e^{-\rho\ts})\ell_0 + 1_{[\ps,\infty )}(\ts )\ell_0 \Big]\\
			&\leq \Ep^{\theta^V}_x\Big[\int_0^{\ts\wedge\ps }e^{-\rho t}\zeta(\hat X^{\theta^V}_t)\rd t + 1_{(0,\ps ]}(\ts )(1-e^{-\rho\ts})\ell_1  + 1_{[\ps,\infty )}(\ts )\ell_1\Big]\\
			&=v_{\ell_1}(x) +\ell_1.
		\end{align*}
		Then we have $x=\xL$ that $ 0=v_{\ell_0}(\xL) +\ell_0\leq v_{\ell_1}(\xL) +\ell_1$. Because $v_{\ell_1}(x)$ increases in $x$, there exists $\delta>0$ such that $v_{\ell_1}(\xL-\delta) +\ell_1 =0$. This means that $\ell\mapsto\xL(\ell)$ is decreasing. Moreover, we have the symmetry assumption from Lemma~\ref{lem:symmetry} that $0=v_{\ell_1}(\xL-\delta) +\ell_1 = -v_{\ell_1}(-\xL+\delta) -\ell_1\leq -v_{\ell_1}(-\xL+\delta) -u.$ Therefore, One can infer the existence of $\delta'>0$ such that $v_{\ell_1}(-\xL+\delta+\delta') +u=0$, implying that $\ell\mapsto\xU(\ell)$ is increasing. This concludes that $(\xL,\xU)$ expand as $\ell$ increases. A similar result holds in the case of increasing $u$, using an analogous argument. It is also clear that the proposition holds when $x>\xS$, after repeating the same procedure. Hence, the proof is complete.
	\end{proof}
	
	When the control cost on one side is higher, the DM responses by delaying the exertion of control on that side, reducing the frequency of action to minimize the running cost. Furthermore, the DM also delays control on the opposite side, in expectation, to prevent the cash level from reaching a threshold where activating the more expensive control becomes necessary. This behavior leads to an expansion of the control barriers, as demonstrated in Proposition 6.
	
	We next explore the effects arising from asymmetries, such as: non-zero drift, unequal holding costs for positive and negative cash balances, and unequal control costs, for different levels of ambiguity.
	
	For simplicity, we suppose from now on that $\kappa_1\geq\kappa_0$, where $\kappa_0\triangleq\kappa$, and that denoted by $v_{\kappa_i}(x)$ the marginal value function \eqref{eq:dynkin} for each $\kappa_i,i=0,1$.
	\begin{proposition}[Comparative statics of ambiguity with non-zero drift] \label{prop:com amb-drift abm}
		Suppose that $\alpha$ is non-zero while the other parameters remain symmetric. Then, the continuation region $(\xL,\xU)$ shrinks as $\kappa$ increases. That is, $\kappa\mapsto\xL(\kappa)$ is non-decreasing and $\kappa\mapsto\xU(\kappa)$ is non-increasing.
	\end{proposition}
	\begin{proof}[Proof of Proposition~\ref{prop:com amb-drift abm}]
		It is easy to see from Proposition~\ref{prop:com amb abm} that \eqref{eq:v kappa} holds, even if $\alpha$ is non-zero. Therefore, it holds that $v_{\kappa_0}(x)\geq v_{\kappa_1}(x)$ for any $x<\xS$. Similarly, it also follows that $v_{\kappa_0}(x)\leq v_{\kappa_1}(x)$ on which $x>\xS$.  This means that there are $\delta,\delta'>0$ such that $v_{\kappa_1}(\xL+\delta)=-\ell$ and $v_{\kappa_1}(\xU-\delta')=u$.
		Hence, we conclude that $(\xL,\xU)$ shrink as $\kappa$ increases for any non-zero drift.
	\end{proof}
	
	With a similar reasoning to the proof of Proposition~\ref{prop:com amb-drift abm}, we obtain the following proposition.
	\begin{proposition}[Comparative statics of ambiguity with unequal holding costs] \label{prop:com amb-holding-cost abm}
		Suppose that $\cc\neq\ch$, while the other parameters remain symmetric. Then, the control barriers $(\xL,\xU)$ shrink as $\kappa$ increases. That is, $\kappa\mapsto\xL(\kappa)$ is non-decreasing and $\kappa\mapsto\xU(\kappa)$ is non-increasing. 
	\end{proposition}
	Propositions~\ref{prop:com amb-drift abm} and~\ref{prop:com amb-holding-cost abm} imply that the ``shrink'' effect caused by increasing ambiguity persists even in the presence of non-zero drift or unequal holding costs for positive and negative cash balances. However, this effect does not hold when unequal control costs are considered, resulting in a non-monotonicity established in the following proposition.
	\begin{proposition}[Comparative statics of ambiguity with unequal control costs] \label{prop:com amb-control-cost abm}
		Suppose that $\ell\neq u$, while the other parameters remain symmetric. Then, the control barriers $(\xL,\xU)$ shrink as $\kappa$ increases if $|\ell-u|$ small enough. Otherwise, there exist $\epsilon>0$ such that $(\xL,\xU)$ shift downward if $\ell-u<\epsilon$, or shift upward if $\ell-u>\epsilon$, as $\kappa$ increases.
	\end{proposition}
	\begin{proof}[Proof of Proposition~\ref{prop:com amb-control-cost abm}]
		Suppose that $\ell = u + \epsilon$ for some $\epsilon > 0$. By Lemma~\ref{lem:symmetry}, we have $\ell = -v_{\kappa_0}(\xL) = -v_{\kappa_1}(\xL + \delta) = -v_{\kappa_1}(-\xL - \delta) > \ell - \epsilon = u$ for some $\delta > 0$. Proposition~\ref{prop:com control cost abm} then implies that there exists $\delta' > 0$ such that $-v_{\kappa_1}(-\xL - \delta + \delta') = u$. It is straightforward to observe that the mapping $\epsilon \mapsto \delta'(\epsilon)$ is increasing with $\delta'(0) = 0$. Consequently, there exist $\underline{\epsilon}, \overline{\epsilon} > 0$ such that $\overline{\epsilon} > \underline{\epsilon}$ and $\delta'(\underline{\epsilon}) < \delta < \delta'(\overline{\epsilon})$. This implies that if $u - \ell$ is sufficiently small, i.e., $\epsilon = \underline{\epsilon}$, then $(\xL, \xU)$ shrink. Conversely, if $u - \ell$ is sufficiently large, i.e., $\epsilon = \overline{\epsilon}$, the control barriers lead to an upward shift. Similarly, it can be shown that $(\xL, \xU)$ shift downward when $\ell - u$ becomes sufficiently large.
	\end{proof}
	
\begin{figure}[!htb]
		\begin{subfigure}{0.45\textwidth}
			\centering
			\includegraphics[width=\linewidth]{ 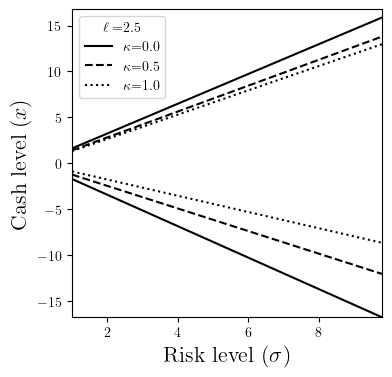}
			\caption{Control barriers given $\ell=2.5$ for $\kappa\in\{0,0.5,1\}$.}\label{fig:CB_L2}
		\end{subfigure}\hfill
		\begin{subfigure}{0.45\textwidth}
			\centering
			\includegraphics[width=\linewidth]{ 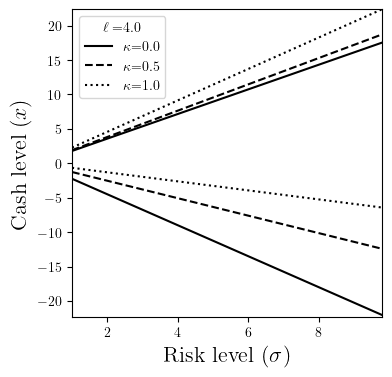}
			\caption{Control barriers given $\ell=4$ for $\kappa\in\{0,0.5,1\}$.}\label{fig:CB_L3}
		\end{subfigure}
		\caption{Control barriers as a function of $\sigma$ for $\kappa\in\{0,0.5,1\}$, with fixed parameters $\alpha=0$, $\rho=0.1$, $u=2$, $\cc=\ch=1$. Panels (\subref{fig:CB_L2}) and (\subref{fig:CB_L3}) displays the control barriers for $\ell=2.5$ and $\ell=4$, respectively.}
	\end{figure}

	This result indicates that when the difference between upper and lower control costs becomes too large, the ``shrink effect'' from ambiguity no longer holds. As shown in Figures~\ref{fig:CB_sym},~\ref{fig:CB_L2} and~\ref{fig:CB_L3}, when the cost of controlling the lower inventory becomes significantly higher than that of the upper, the upper barriers shift upward, contrary to the conventional expectation that more ambiguous DM acts earlier. This occurs due to the mitigation effect of singular control. In the symmetric case, the DM exerts control earlier at either the upper or lower barrier. However, when the cost of controlling one barrier is significantly lower than the other, the optimal policy suggests delaying the cheaper action to reduce the likelihood of triggering the more costly control. In other words, there exists a scenario where the optimal singular control policy offsets the impact of ambiguity, illustrating the duality between singular control and ambiguity.
	
	\begin{figure}[!htb]
		\begin{subfigure}{0.65\textwidth}
			\centering
			\includegraphics[width=\linewidth]{ 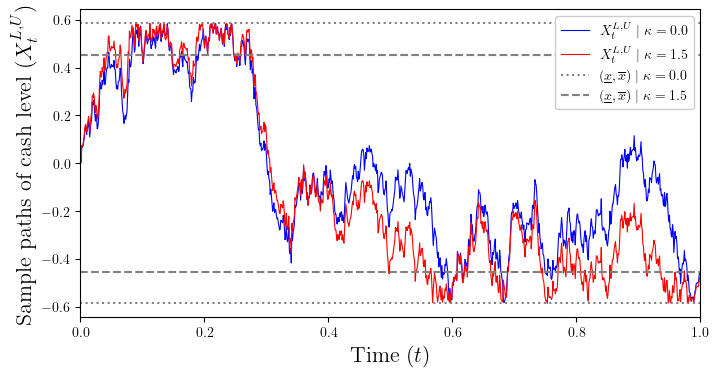}
			\caption{Sample paths}\label{fig:sample_paths}
		\end{subfigure}\hfill
		\begin{subfigure}{0.35\textwidth}
			\centering
			\includegraphics[width=\linewidth]{ 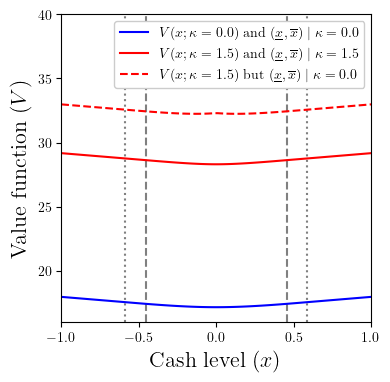}
			\caption{Value functions}\label{fig:vf_ignore}
		\end{subfigure}
		\caption{Panel (\subref{fig:sample_paths}) shows sample paths of cash levels without ambiguity ($\kappa=0$, blue line) and with ambiguity, ($\kappa=1.5$, red line), using base-case parameters: $\alpha = 0$, $\rho = 0.1$, $\sigma=0.5$, $\ell = u = 2$, and $\cc = \ch = 1$. The value $X_0=0$ is fixed for both cases. The dotted and dashed gray lines correspond to the control barriers for $\kappa=0$ and $\kappa=1.5$, respectively. The associated value functions are displayed in Panel (\subref{fig:vf_ignore}), where the solid blue and red lines are for the cases $\kappa=0$ and $\kappa=1.5$, respectively. The dashed red line is the cost function when the worse-case happens $(\kappa=1.5)$, but the DM uses the control policy belonging to that of reference prior $(\kappa=0)$.}  
	\end{figure}
	
	\section{Managerial implication}
	
	According to the earlier argument, firms operating under the model with a presumably known prior tend to incur lower average costs for holding cash. This naturally raises a question: why should ambiguity be taken into account if doing so results in higher operational costs? From the standpoint of revenue optimization, relying on the classical model may initially seem more attractive, as it appears to minimize control and holding expenses.
	
	However, this conclusion only holds if the assumed probability distribution accurately reflects the environment the firm faces. The key managerial insight emerges when a firm disregards ambiguity but then encounters unfavorable outcomes. In such situations, the cost of ignoring uncertainty can become significantly higher than anticipated. For example, consider Figure~\ref{fig:sample_paths}. The blue line represents the standard model, with dotted control barriers indicating the optimal strategy under perceived known risk. In contrast, the red line shows the cash trajectory under the worst-case scenario, guided by dashed control barriers that account for ambiguity. These paths reveal that, under the worst-case prior, the firm's cash position tends to shift more rapidly, rising when the marginal cost of holding cash is high ($X_t>x^\ast$) and falling when it is low ($X_t\leq x^\ast$).
	
	If the firm continues to rely on the standard model while facing a worst-case scenario, it will respond suboptimally. This leads to more frequent interventions and higher holding costs, as illustrated in the top-left and bottom-right panels of Figure~\ref{fig:sample_paths}. In other words, while the standard model may appear cost-effective ex ante, as indicated by the solid blue line of Figure~\ref{fig:vf_ignore}, it can lead to greater losses when uncertainty proves more severe than expected, as depicted by the dashed red line in the same figure. By contrast, adopting an ambiguity-aware approach, such as the maxmin utility model, allows the firm to anticipate such a complication and respond more proactively. Although this requires earlier or more frequent interventions, as shown by the higher value function represented by the solid red line in Figure~\ref{fig:vf_ignore}, it acts as a form of precautionary adjustment. Importantly, the average running cost under this approach is lower than that of the standard model that ignores ambiguity. One can interpret this surplus as an \emph{ambiguity premium}, a strategic cost paid upfront to mitigate future downside uncertainty.
	
	This highlights a critical managerial implication: when ambiguity exerts a first-order effect on cash management, it is no longer a secondary consideration but a central determinant of optimal policy. Firms that incorporate ambiguity into their decision-making frameworks are better positioned to protect firm value in uncertain or poorly understood environments.
	
	\section{Conclusion}
	
	In this paper, we revisit the classical problem of two-sided singular control in the optimal cash reserve problem where the net cash position evolves according to an It\={o} diffusion. We extend this model to allow for managerial ambiguity within the multiple prior framework with $\kappa$-ignorance and maxmin expected utility. We establish a verification theorem for the minimal expected present value for holding and control costs under the worst-case prior, as well as an optimal cash holding policy. A Dynkin game approach is used to explore the effects of the level of ambiguity (and other parameters) on the optimal control of cash holdings.
	
	From a managerial perspective, our most important observation  is that ambiguity increases the frequency with which control is exerted. This is due to the fact that under the worst-case prior the manager expects higher holding costs, which makes exerting control relatively cheaper. This results, on average, in a smaller cash inventory, which is the opposite effect to an increase in risk. If risk, as measured by the variance of the  net cash flow, increases, then the standard ``option value of waiting'' (cf., \citealp{DiPi94}) increases which implies that, typically, control is exerted later. This results, in expectation, in a larger cash inventory.
	
	Our work suggests several avenues for future research.
	First, we note that our findings are purely theoretical, and the numerical results presented serve only as illustrative examples. However, because our model is both analytically tractable and robust for any range of model parameters, comparable to the seminal work of \citet{Nishimura2007-it}, it can readily be used as a benchmark for empirical validation using real-world economic data. We believe that this would be a valuable avenue for future research.
	
	Secondly, the assumption of proportional control costs, such as for equity issuance or dividend payments, may not fully reflect the complexity of real-world cost structures, which could involve both variable and fixed components. The existence of such fixed costs leads to a problem of \emph{impulse control}. For an overview of the method, see, e.g., \citet[Chapter 7]{Michael_Harrison2013-fe}.
	
	Thirdly, one of the assumptions of our model is that the manager does not learn about the set of priors. It is as if the manager is confronted with a new Ellsberg urn at every point in time. In some real-world situations it may be more realistic to assume that the manager is confronted with the \emph{same} Ellsberg urn at every point in time. This then opens up the possibility of managerial learning about ambiguity. 
	
	Finally, our model of $\kappa$-ambiguity describes a fairly extreme version of cautious behavior. A more realistic version of the model would allow the manager to average over multiple priors. That would naturally lead to the smooth ambiguity model of \citet{KlMaMu05}. 
	

	\section*{Acknowledgments}
	We sincerely thank the anonymous referees for their constructive comments, which significantly improved our manuscript. Part of the research was conducted when Hellmann was at the Center for Mathematical Economics (IMW) at Bielefeld University, Germany. Thijssen gratefully acknowledges support from the Center for Mathematical Economics (IMW) and the Center for Interdisciplinary Research (ZiF) at Bielefeld University. Helpful comments were received from participants of the ZiF programm ``Robust Finance: Strategic Power, Knightian Uncertainty, and the Foundations of Economic Policy Advice''. The work of Giorgio Ferrari was funded by the Deutsche Forschungsgemeinschaft (DFG, German Research Foundation), Project ID: 317210226--SFB 1283.
	
	\bibliography{ref}
	
	\newpage
	\appendix
	\appendixpage
	\addappheadtotoc
	
	\section{Proof of Proposition \ref{th:verification2}}\label{pf:verification2}
	
	First note that the constants $A$ and $B$ depend on $\xL$, whereas the constants $C$ and $D$ depend on $\xU$. In what follows we will make this dependence explicit by writing, say, the constant $A$, as a mapping $\xL\mapsto A(\xL)$.  In fact, for given $\xL$ and $\xU$, the systems of linear equations
	\begin{align}\label{eq:lower_coeff}
		\begin{bmatrix}
			\vH'_{-\kappa}(\xL+) & \vC'_{-\kappa}(\xL+)\\
			\vH''_{-\kappa}(\xL+) & \vC''_{-\kappa}(\xL+)
		\end{bmatrix}
		\begin{bmatrix}
			A(\xL)\\B(\xL)
		\end{bmatrix}=
		\begin{bmatrix}
			-\ell-R_{-\kappa}'(\xL)\\-R_{-\kappa}''(\xL)
		\end{bmatrix},
	\end{align}
	and
	\begin{align}\label{eq:upper_coeff}
		\begin{bmatrix}
			\vH'_{+\kappa}(\xU-) & \vC'_{+\kappa}(\xU-)\\
			\vH''_{+\kappa}(\xU-) & \vC''_{+\kappa}(\xU-)
		\end{bmatrix}
		\begin{bmatrix}
			C(\xU)\\D(\xU)
		\end{bmatrix}=
		\begin{bmatrix}
			u-R_{+\kappa}'(\xU)\\-R_{+\kappa}''(\xU)
		\end{bmatrix},
	\end{align}
	have unique solutions: 
	\begin{align}
		A(\xL)&=\frac{\vC''_{-\kappa}(\xL+)(-\ell-R'_{-\kappa}(\xL)) + \vC'_{-\kappa}(\xL+)R''_{-\kappa}(\xL)}{\vC''_{-\kappa}(\xL+)\vH'_{-\kappa}(\xL+)-\vC'_{-\kappa}(\xL+)\vH''_{-\kappa}(\xL+)}, \label{A} \\
		B(\xL) &=\frac{-\vH''_{-\kappa}(\xL+)(-\ell-R'_{-\kappa}(\xL)) - \vH'_{-\kappa}(\xL+)R''_{-\kappa}(\xL)}{\vC''_{-\kappa}(\xL+)\vH'_{-\kappa}(\xL+)-\vC'_{-\kappa}(\xL+)\vH''_{-\kappa}(\xL+)}, \label{B} \\
		C(\xU)&=\frac{\vC''_{+\kappa}(\xU-)(u-R'_{+\kappa}(\xU)) + \vC'_{+\kappa}(\xU-)R''_{+\kappa}(\xU)}{\vC''_{+\kappa}(\xU-)\vH'_{+\kappa}(\xU-)-\vC'_{+\kappa}(\xU-)\vH''_{+\kappa}(\xU-)}, \label{C} \\
		D(\xU)&=\frac{-\vH''_{+\kappa}(\xU-)(u-R'_{+\kappa}(\xU)) - \vH'_{+\kappa}(\xU-)R''_{+\kappa}(\xU)}{\vC''_{+\kappa}(\xU-)\vH'_{+\kappa}(\xU-)-\vC'_{+\kappa}(\xU-)\vH''_{+\kappa}(\xU-)}. \label{D}
	\end{align}
	
	Define the function $\varphi$ on $(\xL,\xU)$ as follows:
	\begin{equation}\label{eq:varphi}
		\varphi(x) =  
		\begin{cases}
			R_{-\kappa}(x)+A(\xL)\vH_{-\kappa}(x)+B(\xL)\vC_{-\kappa}(x) &\text{if $\xL<x<x^\ast$}\\
			R_{+\kappa}(x)+C(\xU)\vH_{+\kappa}(x)+D(\xU)\vC_{+\kappa}(x) &\text{if $x^\ast\leq x<\xU$.}
		\end{cases}
	\end{equation} 
	We establish that $\varphi$ is convex on $(\xL,\xU)$. 
	
	First, suppose that $x\in(\xL,x^*)$ and that $ x^*<0 $. It then holds that
	\begin{align*}
		&\varphi''(x)= R''_{-\kappa}(x) + A(\xL)\vH''_{-\kappa}(x) + B(\xL)\vC''_{-\kappa}(x) \\
		&=R''_{-\kappa}(x) + \frac{\vC''_{-\kappa}(\xL+)(-\ell-R'_{-\kappa}(\xL)) + \vC'_{-\kappa}(\xL+)R''_{-\kappa}(\xL)}{\vC''_{-\kappa}(\xL+)\vH'_{-\kappa}(\xL+)-\vC'_{-\kappa}(\xL+)\vH''_{-\kappa}(\xL+)}\vH''_{-\kappa}(x) \\ & \quad + \frac{-\vH''_{-\kappa}(\xL+)(-\ell-R'_{-\kappa}(\xL)) - \vH'_{-\kappa}(\xL+)R''_{-\kappa}(\xL)}{\vC''_{-\kappa}(\xL+)\vH'_{-\kappa}(\xL+)-\vC'_{-\kappa}(\xL+)\vH''_{-\kappa}(\xL+)}\vC''_{-\kappa}(x) \\
		&=\check{c}f_{-\kappa}''(x) + \hat{E}_{-\kappa}\vH''_{-\kappa}(x)  \\
		& \quad + \frac{\vC''_{-\kappa}(\xL+)(-\ell-\check{c}f'_{-\kappa}(x)(\xL) - \hat{E}_{-\kappa}\vH'_{-\kappa}(\xL+)) + \vC'_{-\kappa}(\xL+)(\check{c}f''_{-\kappa}(x)(\xL) + \hat{E}_{-\kappa}\vH''_{-\kappa}(\xL+))}{\vC''_{-\kappa}(\xL+)\vH'_{-\kappa}(\xL+)-\vC'_{-\kappa}(\xL+)\vH''_{-\kappa}(\xL+)}\vH''_{-\kappa}(x) \\ & \quad - \frac{\vH''_{-\kappa}(\xL+)(-\ell-f_{-\kappa}'(\xL) - \hat{E}_{-\kappa}\vH'_{-\kappa}(\xL+)) + \vH'_{-\kappa}(\xL+)(\check{c}f''_{-\kappa}(x)(\xL) + \hat{E}_{-\kappa}\vH''_{-\kappa}(\xL+))}{\vC''_{-\kappa}(\xL+)\vH'_{-\kappa}(\xL+)-\vC'_{-\kappa}(\xL+)\vH''_{-\kappa}(\xL+)}\vC''_{-\kappa}(x) \\
		&= \left[ \frac{\vC''_{-\kappa}(\xL+)(-\ell-\check{c}f'_{-\kappa}(x)(\xL)) }{ \vC''_{-\kappa}(\xL+)\vH'_{-\kappa}(\xL+)-\vC'_{-\kappa}(\xL+)\vH''_{-\kappa}(\xL+)} \right] \vH''_{-\kappa}(x)	\\ 
		& \quad - \left[ \frac{\vH''_{-\kappa}(\xL+)(-\ell-\check{c}f'_{-\kappa}(x)(\xL))}{\vC''_{-\kappa}(\xL+)\vH'_{-\kappa}(\xL+)-\vC'_{-\kappa}(\xL+)\vH''_{-\kappa}(\xL+)}\right]\vC''_{-\kappa}(x)
		\\
		&= \left[ \frac{(-\ell-\check{c}f'_{-\kappa}(x)'(\xL)) }{ \vC''_{-\kappa}(\xL+)\vH'_{-\kappa}(\xL+)-\vC'_{-\kappa}(\xL+)\vH''_{-\kappa}(\xL+)} \right] ( \vC''_{-\kappa}(\xL+) \vH''_{-\kappa}(x) - \vH''_{-\kappa}(\xL+) \vC''_{-\kappa}(x) )
		\geq 0.
	\end{align*}
	
	The final inequality follows from (i) assumption \eqref{eq:cost_condition}; (ii) strict positivity of $\vC''{-\kappa}(\xL+)\vH'{-\kappa}(\xL+)-\vC'{-\kappa}(\xL+)\vH''{-\kappa}(\xL+)$; and (iii) non-negativity of $\vC''{-\kappa}(\xL+)\vH''{-\kappa}(x)-\vH''{-\kappa}(\xL+)\vC''{-\kappa}(x)$ on $(\xL,x^\ast)$—this term is zero at $\xL+$ and rises on that interval.
	
	When $\xS\ge0$, the same reasoning applies on $(\xL,0)$, so only the convexity of $\varphi$ on $[0,\xS)$ remains. Here the sign of $A(\xL)$ matters (unlike before), while $B(\xL)\le0$ always, prompting separate treatments of $A(\xL)\ge0$ and $A(\xL)<0$. On $[0,x^\ast)$ we use $R_{-\kappa}(x)=\ch f_{-\kappa}(x)+\check{E}\vC_{-\kappa}(x)$.
	
	\noindent\textbf{Case 1}: when $ A(\xL) \geq 0 $, it holds that
	\begin{align*}
		\varphi''(x)&= R''_{-\kappa}(x) + A(\xL)\vH''_{-\kappa}(x) + B(\xL)\vC''_{-\kappa}(x)\\
		&=\hat{c}f''_{-\kappa}(x) + \check{E}_{-\kappa}\vC''_{-\kappa}(x) + A(\xL)\vH''_{-\kappa}(0)\frac{\vH''_{-\kappa}(x)}{\vH''_{-\kappa}(0)} + B(\xL)\vC''_{-\kappa}(0)\frac{\vC''_{-\kappa}(x)}{\vC''_{-\kappa}(0)}\\
		&\geq  \check{E}_{-\kappa}\vC''_{-\kappa}(0)\frac{\vC''_{-\kappa}(x)}{\vC''_{-\kappa}(0)} + A(\xL)\vH''_{-\kappa}(0)\frac{\vC''_{-\kappa}(x)}{\vC''_{-\kappa}(0)} + B(\xL)\vC''_{-\kappa}(0)\frac{\vC''_{-\kappa}(x)}{\vC''_{-\kappa}(0)}	\\
		&= \hat{E}_{-\kappa}\vH''_{-\kappa}(0)\frac{\vC''_{-\kappa}(x)}{\vC''_{-\kappa}(0)}  \\
		& \qquad + \frac{\vC''_{-\kappa}(\xL+)(\ell-\check{c}f'_{-\kappa}(\xL) - \hat{E}_{-\kappa}\vH'_{-\kappa}(\xL+)) + \vC'_{-\kappa}(\xL+)  \hat{E}_{-\kappa}\vH''_{-\kappa}(\xL+)}{\vC''_{-\kappa}(\xL+)\vH'_{-\kappa}(\xL+)-\vC'_{-\kappa}(\xL+)\vH''_{-\kappa}(\xL+)}\vH''_{-\kappa}(0)\frac{\vC''_{-\kappa}(x)}{\vC''_{-\kappa}(0)} \\ 
		& \qquad - \frac{\vH''_{-\kappa}(\xL+)(\ell-\check{c}f'_{-\kappa}(\xL) - \hat{E}_{-\kappa}\vH'_{-\kappa}(\xL+)) + \vH'_{-\kappa}(\xL+) \hat{E}_{-\kappa}\vH''_{-\kappa}(\xL+)}{\vC''_{-\kappa}(\xL+)\vH'_{-\kappa}(\xL+)-\vC'_{-\kappa}(\xL+)\vH''_{-\kappa}(\xL+)}\vC''_{-\kappa}(0)\frac{\vC''_{-\kappa}(x)}{\vC''_{-\kappa}(0)} \\
		&=\hat{E}_{-\kappa}\vH''_{-\kappa}(0)\frac{\vC''_{-\kappa}(x)}{\vC''_{-\kappa}(0)} \\
		& \qquad+\left[ \frac{\vC''_{-\kappa}(\xL+)\vH''_{-\kappa}(0)(\ell-\check{c}f'_{-\kappa}(\xL)) }{ \vC''_{-\kappa}(\xL+)\vH'_{-\kappa}(\xL+)-\vC'_{-\kappa}(\xL+)\vH''_{-\kappa}(\xL+)} - \hat{E}_{-\kappa}\vH''_{-\kappa}(0) \right] \frac{\vC''_{-\kappa}(x)}{\vC''_{-\kappa}(0)} \\ 
		& \qquad - \left[\frac{\vH''_{-\kappa}(\xL+)\vC''_{-\kappa}(0)(\ell-\check{c}f'_{-\kappa}(\xL))}{\vC''_{-\kappa}(\xL+)\vH'_{-\kappa}(\xL+)-\vC'_{-\kappa}(\xL+)\vH''_{-\kappa}(\xL+)}\right] \frac{\vC''_{-\kappa}(x)}{\vC''_{-\kappa}(0)}\\
		&= \left[ \frac{\vC''_{-\kappa}(\xL+)\vH''_{-\kappa}(0)-\vH''_{-\kappa}(\xL+)\vC''_{-\kappa}(0) }{ \vC''_{-\kappa}(\xL+)\vH'_{-\kappa}(\xL+)-\vC'_{-\kappa}(\xL+)\vH''_{-\kappa}(\xL+)} (\ell-\check{c}f'_{-\kappa}(\xL)) \right] \frac{\vC''_{-\kappa}(x)}{\vC''_{-\kappa}(0)}
		\geq 0.
	\end{align*}
	
	\noindent The first inequality holds since $\frac{\vH''_{-\kappa}(x)}{\vH''_{-\kappa}(0)} \geq \frac{\vC''_{-\kappa}(x)}{\vC''_{-\kappa}(0)}$ for all $ x\geq0 $. The last equality obtains from the the fact that $f'_{-\kappa}$ is constant, together with~\eqref{eq:cost_condition}, i.e. $ \hat{E}_{-\kappa}\vH''_{-\kappa}(0) =  \check{E}_{-\kappa}\vC''_{-\kappa}(0)$. Condition~\eqref{eq:cost_condition} and the fact that $ \vC''_{-\kappa}(\xL+)\vH''_{-\kappa}(0)-\vH''_{-\kappa}(\xL+)\vC''_{-\kappa}(0) \geq 0$ give the last inequality. 
	
	\textbf{Case 2}: when $ A(\xL) < 0 $, it holds that
	\begin{align*}
		\varphi''(x)&= R''_{-\kappa}(x) + A(\xL)\vH''_{-\kappa}(x) + B(\xL)\vC''_{-\kappa}(x)\\
		&=\hat{c}f_{-\kappa}''(x) + \check{E}_{-\kappa}\vC''_{-\kappa}(x) + A(\xL)\vH''_{-\kappa}(\xS)\frac{\vH''_{-\kappa}(x)}{\vH''_{-\kappa}(\xS)} + B(\xL)\vC''_{-\kappa}(\xS)\frac{\vC''_{-\kappa}(x)}{\vC''_{-\kappa}(\xS)}\\
		&\geq  \check{E}_{-\kappa}\vC''_{-\kappa}(\xS)\frac{\vC''_{-\kappa}(x)}{\vC''_{-\kappa}(\xS)}   + A(\xL)\vH''_{-\kappa}(\xS)\frac{\vC''_{-\kappa}(x)}{\vC''_{-\kappa}(\xS)} + B(\xL)\vC''_{-\kappa}(\xS)\frac{\vC''_{-\kappa}(x)}{\vC''_{-\kappa}(\xS)}\\
		&= \left(\check{E}_{-\kappa}\vC''_{-\kappa}(\xS)   + A(\xL)\vH''_{-\kappa}(\xS) + B(\xU)\vC''_{-\kappa}(\xS)\right)\frac{\vC''_{-\kappa}(x)}{\vC''_{-\kappa}(\xS)}\\
		&=\left(\check{E}_{+\kappa}\vC''_{+\kappa}(\xS)   + C(\xU)\vH''_{+\kappa}(\xS) + D(\xU)\vC''_{+\kappa}(\xS)\right)\frac{\vC''_{-\kappa}(x)}{\vC''_{-\kappa}(\xS)}\\
		&=\Bigg[\check{E}_{+\kappa}\vC''_{+\kappa}(\xS)   + \frac{\vC''_{+\kappa}(\xU-)(u-\hat{c}f'_{+ \kappa}(\xU) - \check{E}_{+\kappa}\vC'_{+ \kappa}(\xU)) + \vC'_{+\kappa}(\xU-)\check{E}_{+\kappa}\vC''_{+ \kappa}(\xU)}{\vC''_{+\kappa}(\xU-)\vH'_{+\kappa}(\xU-)-\vC'_{+\kappa}(\xU-)\vH''_{+\kappa}(\xU-)}\vH''_{+\kappa}(\xS)  \\  
		&\qquad -\frac{\vH''_{+\kappa}(\xU-)(u-\hat{c}f'_{+ \kappa}(\xU) - \check{E}_{+\kappa}\vC'_{+ \kappa}(\xU)) + \vH'_{+\kappa}(\xU-)\check{E}_{+\kappa}\vC''_{+ \kappa}(\xU)}{\vC''_{+\kappa}(\xU-)\vH'_{+\kappa}(\xU-)-\vC'_{+\kappa}(\xU-)\vH''_{+\kappa}(\xU-)}\vC''_{+\kappa}(\xS)\Bigg]\frac{\vC''_{-\kappa}(x)}{\vC''_{-\kappa}(\xS)}\\ &=\Bigg[\check{E}_{+\kappa}\vC''_{+\kappa}(\xS)   + \frac{\vC''_{+\kappa}(\xU-)(u-\hat{c}f'_{+ \kappa}(\xU) )}{\vC''_{+\kappa}(\xU-)\vH'_{+\kappa}(\xU-)-\vC'_{+\kappa}(\xU-)\vH''_{+\kappa}(\xU-)}\vH''_{+\kappa}(\xS)  \\  
		&\qquad -\left(\frac{\vH''_{+\kappa}(\xU-)(u-\hat{c}f'_{+ \kappa}(\xU) )  }{\vC''_{+\kappa}(\xU-)\vH'_{+\kappa}(\xU-)-\vC'_{+\kappa}(\xU-)\vH''_{+\kappa}(\xU-)} + \check{E}_{+\kappa} \right)\vC''_{+\kappa}(\xS)\Bigg]\frac{\vC''_{-\kappa}(x)}{\vC''_{-\kappa}(\xS)}\\ 
		&= \Bigg[ \frac{  \vC''_{+\kappa}(\xU-)\vH''_{+\kappa}(\xS) -  \vH''_{+\kappa}(\xU-)\vC''_{+\kappa}(\xS)     }{\vC''_{+\kappa}(\xU-)\vH'_{+\kappa}(\xU-)-\vC'_{+\kappa}(\xU-)\vH''_{+\kappa}(\xU-)}(u-\hat{c}f'_{+ \kappa}(\xU))  \Bigg]\frac{\vC''_{-\kappa}(x)}{\vC''_{-\kappa}(\xS)}
		\geq 0.
	\end{align*}
	The first inequality holds since $\frac{\vH''{-\kappa}(x)}{\vH''{-\kappa}(\xS)} \leq \frac{\vC''{-\kappa}(x)}{\vC''{-\kappa}(\xS)}$ for all $x \geq 0$, and the fourth equality follows from~\eqref{eq:SP_x*}. Moreover, the inequality $\vC''{+\kappa}(\xU-)\vH''{+\kappa}(\xS) - \vH''{+\kappa}(\xU-)\vC''{+\kappa}(\xS) \geq 0$, together with~\eqref{eq:cost_condition}, implies that $\varphi'' \geq 0$ in this case as well.
	
	Therefore, $\varphi$ is convex on $(\xL, \xS)$, and by a similar argument, also convex on $[\xS, \xU)$, establishing convexity on the entire interval $(\xL, \xU)$.
	Consequently, $\varphi$ is decreasing on $(\xL, x^\ast)$ and increasing on $(x^\ast, \xU)$. Direct verification confirms that Condition1 of Theorem\ref{th:verification} holds. Condition 2 holds by assumption, while Condition 3 follows from the choice of $A(\xL), B(\xL), C(\xU), D(\xU)$ in~\eqref{eq:lower_coeff} and~\eqref{eq:upper_coeff}. Conditions 4 and 5 are ensured by~\eqref{eq:cost_condition}, and transversality (Condition6) is trivially satisfied via\eqref{eq:bounded prob of the controlled X}. Thus, all assumptions of Proposition~\ref{th:verification} are fulfilled, and the result follows.

\end{document}